\documentclass{aa}
\usepackage[T1]{fontenc}

\usepackage{graphicx}
\usepackage{txfonts}
\usepackage{amssymb,amsfonts,amsmath,bm}
\usepackage[utf8]{inputenc}
\usepackage[T1]{fontenc}
\usepackage{multirow}
\usepackage{lineno}
\usepackage{hyperref}
\hypersetup{
 colorlinks = true,
 allcolors = blue
}
\DeclareUnicodeCharacter{00A0}{~}
\DeclareMathAlphabet{\mathitbf}{OML}{cmm}{b}{it}
\newcommand{\vA}{\mathitbf{A}}

\newcommand{\vAp}{\mathitbf{A}_{\mathrm{0}}}
\newcommand{\vB}{\mathitbf{B}}

\newcommand{\Bz}{\mathit{B_z}}

\newcommand{\Bn}{\mathit{B_n}}
\newcommand{\vBj}{\mathitbf{B}_{\rm J}}

\newcommand{\vBp}{\mathitbf{B}_{\mathrm{0}}}

\newcommand{\vJ}{\mathitbf{J}}
\newcommand{\Jz}{\mathit{J_z}}
\newcommand{\funsign}{|\Phi_m|}

\newcommand{\izunsign}{|I_{\mathrm{z}}|}
\newcommand{\Itotunsign}{|I_{\mathrm{tot}}|}
\newcommand{\Etot}{E}
\newcommand{\Epot}{E_{\mathrm{0}}}
\newcommand{\Efree}{E_{\rm F}}

\newcommand{\Ediv}{E_{\rm div}}
\newcommand{\Ejs}{E_{{\rm J},s}}

\newcommand{\Ejns}{E_{{\rm J},ns}}
\newcommand{\Epns}{E_{{\rm 0},ns}}
\newcommand{\Emix}{E_{\rm mix}}
\newcommand{\Hv}{H_{V}}
\newcommand{\Hpj}{H_{{\rm PJ}}}
\newcommand{\Hj}{H_{{\rm J}}}
\newcommand{\thetaj}{\theta_J}
\newcommand{\tjavg}{\langle\thetaj\rangle}
\newcommand{\cwavg}{\langle{\rm CW\,sin}\theta\rangle}
\newcommand{\fiavg}{\langle|f_i|\rangle}
\newcommand{\fdavg}{\langle|f_d|\rangle}
\newcommand{\Edivprime}{E_{\rm div}/E}
\newcommand{\Emixprime}{|E_{\rm mix}|/\Ejs}
\newcommand{\degree}{^\circ}
\usepackage{color}
\usepackage[normalem]{ulem}

\begin{document} 

\title{The effect of 
%\so{spatial resolution} 
{spatial sampling} \\ on magnetic field modeling and helicity computation}
%\subtitle{your subtitle}
\author{%
J. K. Thalmann\inst{\ref{inst1}} \and Manu Gupta\inst{\ref{inst1}} \and A. M. Veronig\inst{\ref{inst1},\ref{inst2}}%
}
\institute{%
University of Graz, Institute of Physics/IGAM, Universit\"atsplatz 5, 8010 Graz, Austria\label{inst1} 
\email{julia.thalmann@uni-graz.at} 
\and Kanzelh\"ohe Observatory for Solar and Environmental Research, University of Graz, Austria\label{inst2}%
}
%\date{}
\date{Accepted by A\&A on April 19, 2022}

% \abstract{}{}{}{}{} 
% 5 {} token are mandatory
 
\abstract
{%\textit{Context:}
Nonlinear force-free (NLFF) modeling is regularly used to indirectly infer the 3D geometry of the coronal magnetic field, which is not otherwise accessible on a regular basis by means of direct measurements. 
} 
{%\textit{Aims:} 
We study the effect of binning in time-series NLFF modeling of individual active regions (ARs) in order to quantify the effect of a different underlying %\so{spatial resolution}
 {spatial sampling} on the quality of modeling as well as on the derived physical parameters.
} 
{%\textit{Methods:} 
We apply an optimization method to sequences of Solar Dynamics Observatory (SDO) Helioseismic and Magnetic Imager (HMI) vector magnetogram data at three different %\so{spatial resolutions} 
{plate scales} for three solar active regions to obtain nine NLFF model %\so{time-series} 
{time series}. From the NLFF models, we deduce active-region magnetic fluxes, electric currents, magnetic energies, and relative helicities, and analyze those with respect to the underlying %\so{spatial resolution} 
{spatial sampling}. We calculate various metrics to quantify the quality of the derived NLFF models and apply a Helmholtz decomposition to characterize solenoidal errors.
} 
{%\textit{Results:} 
At a given %\so{spatial resolution} 
{spatial sampling}, the quality of NLFF modeling is different for different ARs, and the quality varies along the individual model %\so{time-series} 
{time series}. For a given AR, modeling at a %\so{given} 
{certain} %\so{spatial resolution} 
{spatial sampling} is not necessarily of superior quality compared to that performed %\so{at different spatial resolutions} 
{with a different plate scale.} %\so{at all time instances of a NLFF model time-series}.
Generally, the NLFF model quality tends to be higher %\so{at reduced spatial resolution} 
{for larger pixel sizes} with the solenoidal quality being the ultimate cause for systematic variations in model-deduced physical quantities.
}
{%\textit{Conclusions:} 
Optimization-based modeling %\so{using binned SDO/HMI vector data delivers} 
{using SDO/HMI vector data binned to larger pixel sizes yields variations in} magnetic %\so{energies} 
{energy} and helicity estimates %\so{that are different by}
{of} $\lesssim$30\% {on overall},
%\LEt{ different from what? Please consider rewording to clarify this point.} 
given that concise checks ensure the physical plausibility and high solenoidal quality of the tested model. %\so{Spatial-resolution-induced} 
{Spatial-sampling-induced} differences are relatively small compared to those arising from other sources of uncertainty, including the effects of applying different data calibration methods, those of using vector data from different instruments, or those arising from application of different NLFF methods to identical input data.
} 

% keywords

\titlerunning{The effect of %\so{spatial resolution} 
{spatial sampling} on magnetic field modeling and helicity computation}
\authorrunning{Thalmann, J.~K., et al.}
\maketitle
%-------------------------------------------------------------------

%%%%%%%%%%%%%%%%%%%%%%%%%%%%%%%%%%%%%%%%%%%%%%%%%%%%%%%%%%%%%%%%%%%%%%%%%%%%%%%%%%%%%
\section{Introduction}
\label{s:intro}
%%%%%%%%%%%%%%%%%%%%%%%%%%%%%%%%%%%%%%%%%%%%%%%%%%%%%%%%%%%%%%%%%%%%%%%%%%%%%%%%%%%%%

To date, three-dimensional (3D) models of the coronal magnetic field are commonly used to obtain insights into related physical processes \citep[][]{2017SSRv..210..249W}. Corresponding modeling approaches are needed because of the otherwise sparse direct measurements of the coronal magnetic field vector even within limited coronal volumes \citep[e.g. review by][]{2009SSRv..144..413C}. In particular, nonlinear force-free (NLFF) magnetic field models are most often used \citep[for dedicated reviews see, e.g.,][]{2012LRSP....9....5W,2013SoPh..288..481R} which are static approximations of the magnetized coronal plasma being necessarily in equilibrium when the Lorentz force vanishes, that is, when gas pressure and other forces are negligible. These conditions are satisfied to a high degree in the active-region corona \citep[e.g.,][]{2001SoPh..203...71G}.

The computation of a force-free magnetic field, $\vB$, requires the numerical solution of
\begin{equation}
        \left(\nabla\times\vB\right) \times \vB = \bf{0}, \label{eq:ff1} 
\end{equation}
and
\begin{equation}
        \nabla\cdot\vB = 0, \label{eq:ff2}
\end{equation}
within a 3D volume, $V$, subject to conditions specified on the lower boundary of the model volume  at $z=0$. In other words, the magnetic field information at the lower boundary of the model is ``extrapolated'' into the coronal volume above. Ideally, in order to specify suitable boundary conditions to solve Eqs.~(\ref{eq:ff1})--(\ref{eq:ff2}), we would hope to have spectro-polarimetric observations at hand that would allow us to deduce a corresponding magnetic field vector consistent with the force-free assumption, for instance measured at chromospheric heights \citep[e.g.,][]{1995ApJ...439..474M}. However, in practice, such data at high spatial and temporal resolution are obtained only from measurements at photospheric heights. The latter are known to represent a regime inconsistent with the force-free approach because of non-negligible gas pressure and gravitational forces. Force-free modeling carried out on the basis of such inconsistent data is known to result in larger residual Lorentz force and divergence, yet may be partially compensated by, for example, preprocessing of the photospheric vector data prior to extrapolation \citep[e.g.,][]{2006SoPh..233..215W,2011A&A...526A..70F}, allowing the force-free solution to deviate from the actually supplied input data at $z=0$ \citep[e.g.,][]{2010A&A...516A.107W,2009ApJ...700L..88W,2011ApJ...728..112W}, or both \citep[][]{2012SoPh..281...37W}.

In any case, force-free modeling at the full available spatial and temporal scales may be numerically expensive. Thus, to realize the modeling of, for example, active regions (ARs) during their disk passage within a reasonable amount of time, the photospheric vector magnetic field data are often spatially binned prior to their usage. \cite{2015ApJ...811..107D} provided the first comprehensive study of the effect of %\so{spatial resolution} 
{spatial sampling} on NLFF modeling by testing the effect of binning of {\it Hinode}/Solar Optical Telescope (SOT) \citep[][]{2008SoPh..249..167T,2013SoPh..283..579L} Stokes spectra onto the model outcome of five different numerical methods. In particular, they binned the Stokes spectra from a chosen spectral scan of AR~10978 using nine different integer factors. Those nine spectra were then subjected
%\LEt{Please check that I have retained your intended meaning.} 
to spectro-polarimetric inversion, 180$\degree$ ambiguity resolution, and remapping to a planar grid, ultimately representing the input data for subsequent NLFF modeling. Correspondingly, the employed single-snapshot models were at nine different %\so{spatial resolutions with} 
{plate scales} ranging from $\sim$0.1~Mm to $\sim$1.7~Mm.

%%%%%%%%%%%%%%%%%%%%%%%%%%%%%%%%%%%%%%%%%%%%%%%%%%%%%%%%%%%%%%%%%%%%%%%%%%%%%%%%%%%%%
\begin{figure*}[t]
\centering
\includegraphics[width=0.9\textwidth]{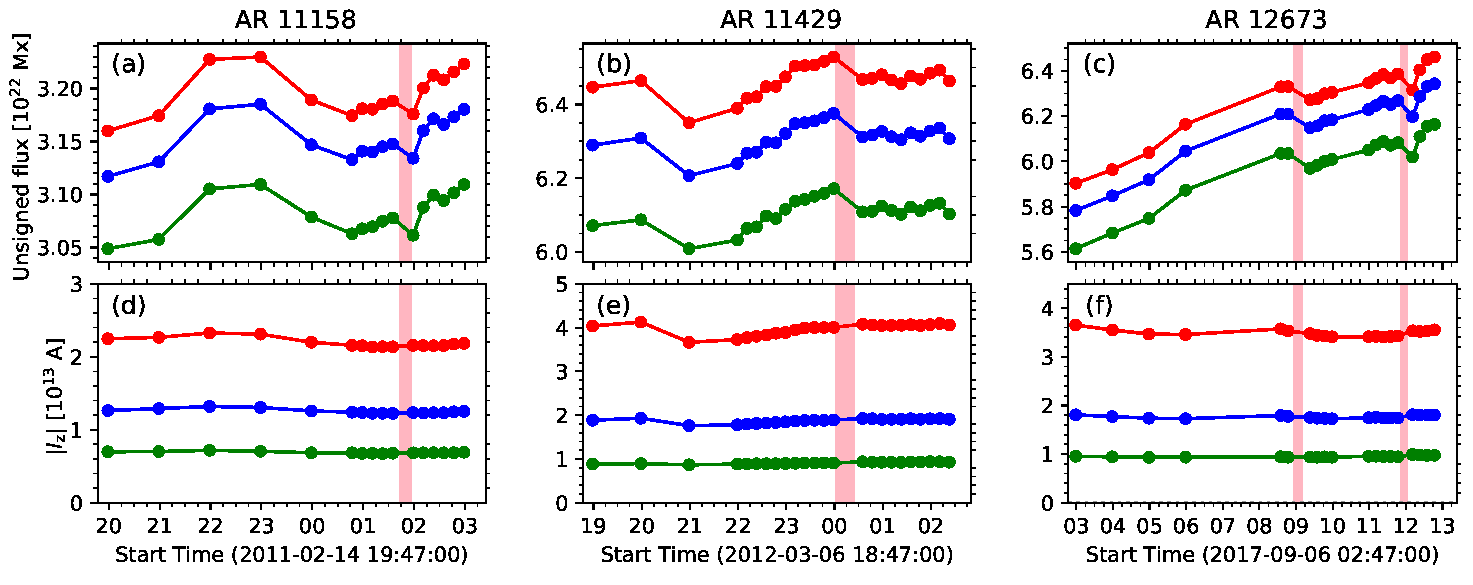}
\caption{Area-integrated parameters computed from the photospheric vector data {at three different pixel sizes} as a function of time for three ARs. 
%\so{, and at three different spatial resolutions}
{\it Top:} Unsigned magnetic flux, $\funsign$. {\it Bottom:} Unsigned vertical current, $\izunsign$. Red, blue, and green colors correspond to bin1, bin2, and bin4 data, respectively, with inherent plate scales of $\sim$0.36, $\sim$0.72, and $\sim$1.4~Mm. Vertical bars indicate the impulsive phase of X-class flares that occurred during the time intervals analyzed.}
\label{fig:fig1}
\end{figure*}
%%%%%%%%%%%%%%%%%%%%%%%%%%%%%%%%%%%%%%%%%%%%%%%%%%%%%%%%%%%%%%%%%%%%%%%%%%%%%%%%%%%%%

In order to obtain a more differentiated picture regarding the effect of 
%\so{spatial resolution} 
{spatial sampling} onto NLFF modeling, we %\so{performed} 
{perform} a corresponding in-depth analysis using one of the numerical methods compared in \cite{2015ApJ...811..107D}, namely the optimization method of \cite{2012SoPh..281...37W}, which has been widely applied within the solar community for the purpose of coronal magnetic field modeling. In contrast to earlier studies, we %\so{did} 
{do} not base our analysis on a single-snapshot NLFF model of a single AR, but instead %\so{used} 
{use} %\so{time-series} 
{time series} of NLFF models during extended periods of time for different ARs (Sect.~\ref{ss:target_ars}). This setting allows us to explore the potential of known metrics  in great detail in order to quantify the quality of NLFF solutions (Sect.~\ref{ss:nlff_quality}), and to explore the effect of %\so{spatial resolution} 
{spatial sampling} on the derived quantities, including magnetic fluxes and currents in 2D as well as energies and helicities in 3D (Sect.~\ref{ss:nlff_phys}). The analysis of magnetic helicity, a quantity characterizing the structural complexity of the magnetic field \citep[e.g.,][]{1969JFM....35..117M}, has recently gained particular attention, as well as its sensitivity to the quality of the underlying magnetic field model \citep[][]{2012SoPh..278..347V,2019ApJ...880L...6T,2020A&A...643A.153T}. Ultimately, the employed sequences of NLFF models for several ARs allow us to deduce and discuss resolution-dependent trends (Sect.~\ref{s:discussion}), and to compare those to other effects known to cause uncertainties in coronal magnetic field modeling (instrumental, data processing, etc.).

%%%%%%%%%%%%%%%%%%%%%%%%%%%%%%%%%%%%%%%%%%%%%%%%%%%%%%%%%%%%%%%%%%%%%%%%%%%%%%%%%%%%%
\section{Data and Methods}
\label{s:data_methods}
%%%%%%%%%%%%%%%%%%%%%%%%%%%%%%%%%%%%%%%%%%%%%%%%%%%%%%%%%%%%%%%%%%%%%%%%%%%%%%%%%%%%%

%%%%%%%%%%%%%%%%%%%%%%%%%%%%%%%%%%%%%%%%%%%%%%%%%%%%%%%%%%%%%%%%%%%%%%%%%%%%%%%%%%%%%
\subsection{Active region selection} \label{ss:target_ars}
%%%%%%%%%%%%%%%%%%%%%%%%%%%%%%%%%%%%%%%%%%%%%%%%%%%%%%%%%%%%%%%%%%%%%%%%%%%%%%%%%%%%%

For our study, we select three out of the ten ARs analyzed in \cite{2021A&A...653A..69G}, namely NOAAs 11158, 11429, and 12673, which hosted the top four solar flares (in terms of peak soft X-ray flux) during solar cycle 24 that occurred within $\pm35^\circ$ of the central meridian (see Table~\ref{tab:target_ars}). The time window for analysis is chosen as in \cite{2021A&A...653A..69G}, that is, it covers a time interval of several hours around the occurrence of the X-class flares, as is the time cadence (a 12min time cadence within $\pm$1h around the flare peak time and a 1h cadence otherwise). Data possibly available during the impulsive phases of the flares were not considered due to the limited validity of the force-free assumption during eruptive processes. Accordingly, the number of considered snapshots is different for each of the target ARs, based on the number of X-class flares within the analysis time window and also affected by the availability photospheric vector magnetic field data needed as an input for the analysis.

%%%%%%%%%%%%%%%%%%%%%%%%%%%%%%%%%%%%%%%%%%%%%%%%%%%%%%%%%%%%%%%%%%%%%%%%%%%%%%%%%%%%%
\begin{table}[t]
\scriptsize
\caption{Properties of active regions under study, including NOAA number, GOES SXR class of X-class flares
that occurred during the analysis time window, and their respective on-disk location, followed by the 
number of vector magnetic field maps used.}    
\label{tab:target_ars}  
\centering       
\begin{tabular}{c c c c c }  
\hline\hline    
NOAA & Flare & Flare & Analysis time window (UT) & No. of \\ 
AR no. & class & location & (YYYY-MM-DD hh:mm -- DD hh:mm) & maps \\ 
\hline      
 11158 & X2.2 & S20W10 & 2011-02-14 19:59 -- 15 02:59\color{white}$^{\rm(a)}$ & 17 \\  
 11429 & X5.4 & N18E31 & 2012-03-06 18:59 -- 07 02:23\color{white}$^{\rm(a)}$ & 26 \\
 \hline
 \multirow{2}{*}{ 12673 } & X2.2 & S08W32 & \multirow{2}{*}{2017-09-06 02:59 -- 06 12:47$^{\rm(a)}$} & \multirow{2}{*}{22}\\
 & X9.3 & S09W34\\
\hline         
\end{tabular}
\tablefoot{$^{\rm(a)}$No SDO/HMI data available between 6~September 06:00~UT and 08:48~UT.}
\end{table}
%%%%%%%%%%%%%%%%%%%%%%%%%%%%%%%%%%%%%%%%%%%%%%%%%%%%%%%%%%%%%%%%%%%%%%%%%%%%%%%%%%%%%

%%%%%%%%%%%%%%%%%%%%%%%%%%%%%%%%%%%%%%%%%%%%%%%%%%%%%%%%%%%%%%%%%%%%%%%%%%%%%%%%%%%%%
\subsection{Vector magnetic field data} \label{ss:data}
%%%%%%%%%%%%%%%%%%%%%%%%%%%%%%%%%%%%%%%%%%%%%%%%%%%%%%%%%%%%%%%%%%%%%%%%%%%%%%%%%%%%%

We use %\so{time-series} 
{time series} of vector magnetic field data as originally prepared by \cite{2021A&A...653A..69G}, who use {\sc hmi.sharp\_CEA\_720s} data within automatically identified active-region patches \citep{2014SoPh..289.3549B} constructed from polarization measurements of the Helioseismic and Magnetic Imager \citep[HMI;][]{2012SoPh..275..207S} on board the Solar Dynamics Observatory \citep[SDO;][]{2012SoPh..275....3P}, and projected onto a (local) heliographic plane \citep[][]{1990SoPh..126...21G}. In addition to using %\so{time-series} 
{time series} of photospheric vector magnetic field data at a native plate scale (0.36~Mm at disk center; hereafter referred to as ``bin1''), we construct corresponding ``bin2'' and ``bin4'' %\so{time-series} 
{time series} by binning the original-resolution data by a factor of 2 and 4, respectively, that is, by adhering to an effective plate scale {(pixel size)}  of 0.72~Mm and 1.44~Mm, respectively. The binning applied to %\so{reduce the} 
{to mimic a reduced spatial} resolution of the data uses nearest-neighbor averaging, that is, taking the magnetic field data of 2$\times$2 (4$\times$4) neighboring pixels and calculating the mean value {\citep[for alternative choices as well as an in-depth study of corresponding effects onto the vector magnetic field data see][]{2012SoPh..277...89L}}.

In order to inspect the effect of binning of the vector magnetic field data (later used as an input for NLFF modeling; see Sect.~\ref{ss:nlff}), we compute two commonly used area-integrated quantities for each of our target ARs at each %\so{spatial resolution} 
{of the plate scales}, namely the total unsigned magnetic flux, $\funsign$, defined as
\begin{linenomath*}
\begin{equation}
\funsign=\int_{S\,(z=0)}|\Bz| \cdot \mathrm{d}S, \label{eq:funsign}
\end{equation}
\end{linenomath*}
with $\Bz$ being the vertical component of the vector magnetic field data, and the unsigned vertical current, $\izunsign$, defined as
\begin{linenomath*}
\begin{equation}
\izunsign=\int_{S\,(z=0)}|\Jz| \cdot \mathrm{d}S, \label{eq:izunsign}
\end{equation} 
\end{linenomath*}
where $J_z$ is the vertical current density and $\mu_0\,J_z=(\nabla\times\vB)_{z=0}$. 

For the bin[2,4]-based estimates, we compute average changes across the %\so{time-series} 
{time series} of the individual ARs with respect to the respective original-resolution(bin1-data)-based estimates as
\begin{linenomath*}
\begin{equation}
        \delta\xi_{{\rm bin}X}=\frac{1}{n_t} \sum_{i=1}^{n_t} \frac{\xi_{{\rm bin}X}(t_i)-\xi_{\rm bin1}(t_i)}{10^{-2} \xi_{\rm bin1}(t_i)}\label{eq:tmean}
,\end{equation} 
\end{linenomath*}
where $X=[2,4]$ for the bin[2,4]-based modeling and $n_t$ is the total number of time instances, $n_t=\sum t_i$. 

The {binned (down-sampled)} HMI data %\so{at successively lower spatial resolutions} 
exhibits successively lesser unsigned fluxes and %\so{current}
{currents}. This is actually expected because the binning necessarily reduces the strength of and gradients within the original magnetic field data. From application of Eq.~(\ref{eq:tmean}) to the %\so{time-series} 
{time series} of unsigned fluxes, we find $\delta\funsign$$\simeq$[$-$1.3$\pm$0.1,$-$3.6$\pm$0.1]\% for the bin[2,4] vector magnetic field data of AR~11158 (Fig.~\ref{fig:fig1}(a)), [$-$2.4$\pm$0.0,$-$5.5$\pm$0.1]\% for AR~11429 (Fig.~\ref{fig:fig1}(b)), and [$-$1.9$\pm$0.1,$-$4.8$\pm$0.2]\% for AR~12673 (Fig.~\ref{fig:fig1}(c)). 

For the unsigned vertical current, we find $\delta\izunsign$$\simeq$[$-$42.8$\pm$0.4,$-$68.6$\pm$0.4]\% for AR~11158, [$-$52.8$\pm$0.3,$-$77.1$\pm$0.4]\% for AR~11429, and [$-$49.2$\pm$1.2,$-$72.6$\pm$0.9]\% for AR~12673 (Fig.~\ref{fig:fig1}(d)--(f), respectively). This is in line with the successively lower electric current found from %\so{lower resolution} 
{down-sampled} data in \cite{2015ApJ...811..107D}. Based on binned SOT Spectro-polarimeter (SP) spectra, relative changes of the mean vertical current density of [$-$43.8,$-$68,6]\% can be deduced for spatial scales corresponding to those used in our study. To compute those percentages, we first defined a reference level from their ``bin3'' and ``bin4'' cases, corresponding to an approximate plate scale of $\sim$0.37~Mm, which is comparable to the plate scale our bin1 case. The corresponding estimates of $\langle\Jz\rangle$ (cf.\ their Fig. 4c) were then used to compute a corresponding average value of $\langle\left[\mathit{J_{z,{\rm bin3}}},\mathit{J_{z,{\rm bin4}}}\right]\rangle$$\simeq$12~mA\,m$^{-2}$. Similarly, we compute $\langle\left[\mathit{J_{z,{\rm bin6}}},\mathit{J_{z,{\rm bin8}}}\right]\rangle$$\simeq$6.8~mA\,m$^{-2}$ for an approximate plate scale of $\sim$0.74~Mm, which is comparable to our bin2 case, and use their their $\mathit{J_{z,{\rm bin14}}}$ at a plate scale of $\sim$1.48~Mm as comparable to our bin4 case.

%%%%%%%%%%%%%%%%%%%%%%%%%%%%%%%%%%%%%%%%%%%%%%%%%%%%%%%%%%%%%%%%%%%%%%%%%%%%%%%%%%%%%
\subsection{Magnetic field modeling} \label{ss:nlff}
%%%%%%%%%%%%%%%%%%%%%%%%%%%%%%%%%%%%%%%%%%%%%%%%%%%%%%%%%%%%%%%%%%%%%%%%%%%%%%%%%%%%%

We employ NLFF models from the data %\so{time-series} 
{time series} at the three different plate scales ($\sim$0.36, $\sim$0.72, and $\sim$1.4~Mm) for each of the three target ARs. We use the method of \cite{2012SoPh..281...37W}, which involves two main computational steps, a preprocessing of the 2D input data \citep[to retrieve a force-free consistent boundary condition at $z=0$][]{2006SoPh..233..215W} and subsequent extrapolation \citep[allowing deviations from the input data at $z=0$ in order to account for measurement uncertainties][]{2010A&A...516A.107W}. During both of these steps, larger freedom is given to changes in the horizontal magnetic field component than to changes in the vertical magnetic field component, in accordance with the generally lower measurement accuracy of the former compared to the latter. For completeness we note here that the trend of %\so{higher resolution} 
data {at smaller pixel sizes} %\so{hosting} 
{to host} more unsigned flux and stronger vertical currents is preserved during preprocessing as well as during optimization. In contrast to \cite{2021A&A...653A..69G}, we employ only one %\so{time-series} 
{time series} per target AR and %\so{spatial resolution} 
{plate scale} using standard model parameter settings. In other words, we omit the tuning of model parameters in order to improve the NLFF model results \citep[for dedicated in-depth studies see, e.g.,][]{2019ApJ...880L...6T,2020A&A...643A.153T}. This is because we want to obtain insights into the effects that are purely attributable to the different %\so{spatial resolution} 
{spatial sampling} of the %\so{(input)}
data used {as input} for NLFF modeling and to avoid complicating the (already complex) interpretation of dependencies. Thus, we compute 195 NLFF models in total (at three different %
%\so{spatial resolutions} 
{plate scales} for the considered number of time instance listed in the last column in Table~\ref{tab:target_ars}).

From each modeled NLFF solution for $\vB$, we compute the unsigned magnetic flux and unsigned vertical current using Eqs.~(\ref{eq:funsign}) and (\ref{eq:izunsign}), respectively, at the lower boundary of the NLFF models  ($z=0)$. In addition, we compute the total (volume-integrated) magnetic energy, $\Etot$, as,
\begin{linenomath*}
\begin{equation}
\Etot=\int_V |\vB|^2 {\rm ~d}V. \label{eq:etot}
\end{equation}
\end{linenomath*}
Correspondingly, we compute the potential energy, $\Epot$, using the current-free (minimum-energy) magnetic field solution, $\vBp$, in Eq.~(\ref{eq:etot}). The latter is defined as $\vBp=\nabla\phi$, with $\phi$ being the scalar potential, which is subject to the constraint $\nabla_n\phi=\Bn$ on the volume-bounding surface, $\partial{V}$. We are therefore also able to compute the free magnetic energy as $\Efree=\Etot-\Epot$.

%%%%%%%%%%%%%%%%%%%%%%%%%%%%%%%%%%%%%%%%%%%%%%%%%%%%%%%%%%%%%%%%%%%%%%%%%%%%%%%%%%%%%
\subsubsection{Quality measures} \label{sss:m_quality}
%%%%%%%%%%%%%%%%%%%%%%%%%%%%%%%%%%%%%%%%%%%%%%%%%%%%%%%%%%%%%%%%%%%%%%%%%%%%%%%%%%%%%

For the purpose of quantifying the force-freeness of the obtained NLFF model magnetic fields in three dimensions, we use the current-weighted angle between the modeled magnetic field and the electric current density, $\thetaj$, related to the otherwise often-used current-weighted average of the sine of the angle between the current density and the magnetic field, $\sigma_J$ \citep[``CW\,sin$\theta$'';][]{2000ApJ...540.1150W} by $\tjavg={\rm sin}^{-1}\sigma_J$. As is common practice, we compute the average angle over all grid points, $\tjavg$. For a completely force-free field, $\tjavg=0$.

In order to determine the degree of solenoidality, we employ several different measures commonly used for such purposes. On the one hand, we use the fractional flux as defined in \cite{2020ApJ...900..136G}, namely $\fdavg=(6\,\delta x)\,\fiavg$, with $\delta x$ representing the spacing of the Cartesian mesh and $\fiavg$ representing the volume-average of the magnitude of the fractional flux increase in a small discrete volume about each grid point \citep{2000ApJ...540.1150W}. Being nearly resolution-invariant, $\fdavg$ serves as an ideal tool for the comparison of the solenoidal levels of NLFF models %\so{at different spatial resolutions} 
{with different inherent pixel sizes} while covering the same physical volume.

On the other hand, we use measures based on the decomposition of the magnetic energy into solenoidal and nonsolenoidal parts, the latter being nonzero if the considered magnetic field is not exactly divergence free. \cite{2013A&A...553A..38V} defined a corresponding measure as $\Edivprime$, quantifying the fraction of the total magnetic energy which is related to the nonzero divergence of a tested 3D field $\vB$, where $\Ediv=\Epns+\Ejns+|\Emix|$. Here, $\Epns$ and $\Ejns$ are the energies of the nonsolenoidal components of the potential and current-carrying ($\vBj=\vB-\vBp$) magnetic field, respectively, and $\Emix$ is a mixed potential-current-carrying term \citep[see Eq.~(8) of][for details]{2013A&A...553A..38V}, the latter usually representing the largest contribution to the nonsolenoidal energies (see Sect.~5 in this latter work and also Sect. 3.1.2 in \cite{2015ApJ...811..107D}). Dedicated follow-up studies examined that $\Edivprime\simeq0.1$ (at the most) is to be tolerated when $\vB$ is used for subsequent computation of magnetic helicity \citep{2016SSRv..201..147V,2019ApJ...880L...6T}. In that context, \cite{2020A&A...643A.153T} suggested using an even more restrictive quantity, namely the ratio $\Emixprime$ as a criterion to disqualify a given $\vB$ for subsequent helicity computation, where $\Ejs$ is the energy of the solenoidal component of the current-carrying field (equivalent to the free magnetic energy in a perfectly solenoidal field), and suggested a corresponding threshold to be respected as $\Emixprime$$\lesssim$0.4. As the purpose of our work is to explore all effects caused by a change in the %\so{spatial resolution} 
{spatial sampling}  of the input data, we check whether or not $\Edivprime$=0.1 and/or $\Emixprime$=0.4 are exceeded in our NLFF model %\so{time-series} 
{time series}, yet do not exclude them from subsequent helicity computation. In order to identify corresponding time instances, we mark them separately in the figures of Section \ref{s:results}. This allows us to understand the %\so{resolution-induced} 
{spatial-sampling-induced} variations to a reliable helicity computation. %\so{in that matter.}

%%%%%%%%%%%%%%%%%%%%%%%%%%%%%%%%%%%%%%%%%%%%%%%%%%%%%%%%%%%%%%%%%%%%%%%%%%%%%%%%%%%%
\subsection{Magnetic helicity and its computation} \label{ss:helicity}
%%%%%%%%%%%%%%%%%%%%%%%%%%%%%%%%%%%%%%%%%%%%%%%%%%%%%%%%%%%%%%%%%%%%%%%%%%%%%%%%%%%%

The gauge-invariant relative magnetic helicity in a volume, $V$, can be written as \citep{1984JFM...147..133B,1984CPPCF...9..111F}
\begin{linenomath*}
\begin{equation}
\Hv=\int_V\left(\vA+\vAp\right)\cdot\left(\vB-\vBp\right) {\rm ~d}V, \label{eq:hv}
\end{equation}
\end{linenomath*}
where $\vA$ and $\vAp$ are the respective vector potentials satisfying $\vB=\nabla\times\vA$ and $\vBp=\nabla\times\vAp$. $\Hv$ in Eq.~(\ref{eq:hv}) can be decomposed as, $\Hv=\Hj+\Hpj$ \citep{1999PPCF...41B.167B, 2003and..book..345B}, with
\begin{linenomath*}
\begin{eqnarray}
\Hj &=& \int_V \left(\vA-\vAp\right)\cdot\left(\vB-\vBp\right) {\rm ~d}V, \label{eq:hj}\\
\Hpj &=& 2\int_V \vAp\cdot\left(\vB-\vBp\right) {\rm ~d}V, \label{eq:hpj}
\end{eqnarray}
\end{linenomath*}
where $\Hj$ is the magnetic helicity of the current-carrying field, $\vBj$, and $\Hpj$ is the volume-threading helicity, both being separately gauge invariant \citep[][]{2018ApJ...865...52L}. 

We compute the vector potentials $\vA$ and $\vAp$ required for the computation of the relative helicities in Eqs.~(\ref{eq:hv})--(\ref{eq:hpj}) using the method of \cite{2011SoPh..272..243T}. The method solves systems of partial differential equations to obtain the vector potentials $\vA$ and $\vAp$ using the Coulomb gauge, $\nabla\cdot\vA=\nabla\cdot\vAp=0$. The method has been shown to provide superior solutions of $\vA$ and $\vAp$  regarding their degree of solenoidality and to deliver helicities in line with those produced using other existing methods \citep[][]{2016SSRv..201..147V}.

%%%%%%%%%%%%%%%%%%%%%%%%%%%%%%%%%%%%%%%%%%%%%%%%%%%%%%%%%%%%%%%%%%%%%%%%%%%%%%%%%%%%%
%%%%%%%%%%%%%%%%%%%%%%%%%%%%%%%%%%%%%%%%%%%%%%%%%%%%%%%%%%%%%%%%%%%%%%%%%%%%%%%%%%%%%
\section{Results}\label{s:results}
%%%%%%%%%%%%%%%%%%%%%%%%%%%%%%%%%%%%%%%%%%%%%%%%%%%%%%%%%%%%%%%%%%%%%%%%%%%%%%%%%%%%%
%%%%%%%%%%%%%%%%%%%%%%%%%%%%%%%%%%%%%%%%%%%%%%%%%%%%%%%%%%%%%%%%%%%%%%%%%%%%%%%%%%%%%

In the following, we summarize the quality (Sect.~\ref{ss:nlff_quality}) of the NLFF modeling as well as deduced physical quantities (Sect.~\ref{ss:nlff_phys}) %\so{at different spatial resolutions}  
{performed on the basis of HMI data at different plate scales}. To do so, we compute time-series-averaged changes with respect to the original-resolution (bin1) model %\so{time-series} 
{time series}, by evaluating Eq.~(\ref{eq:tmean}) for the analyzed quantities.

%%%%%%%%%%%%%%%%%%%%%%%%%%%%%%%%%%%%%%%%%%%%%%%%%%%%%%%%%%%%%%%%%%%%%%%%%%%%%%%%%%%%
\subsection{NLFF model quality} \label{ss:nlff_quality}
%%%%%%%%%%%%%%%%%%%%%%%%%%%%%%%%%%%%%%%%%%%%%%%%%%%%%%%%%%%%%%%%%%%%%%%%%%%%%%%%%%%%

%%%%%%%%%%%%%%%%%%%%%%%%%%%%%%%%%%%%%%%%%%%%%%%%%%%%%%%%%%%%%%%%%%%%%%%%%%%%%%%%%%%%%
\begin{figure*}
\centering
\includegraphics[width=\textwidth]{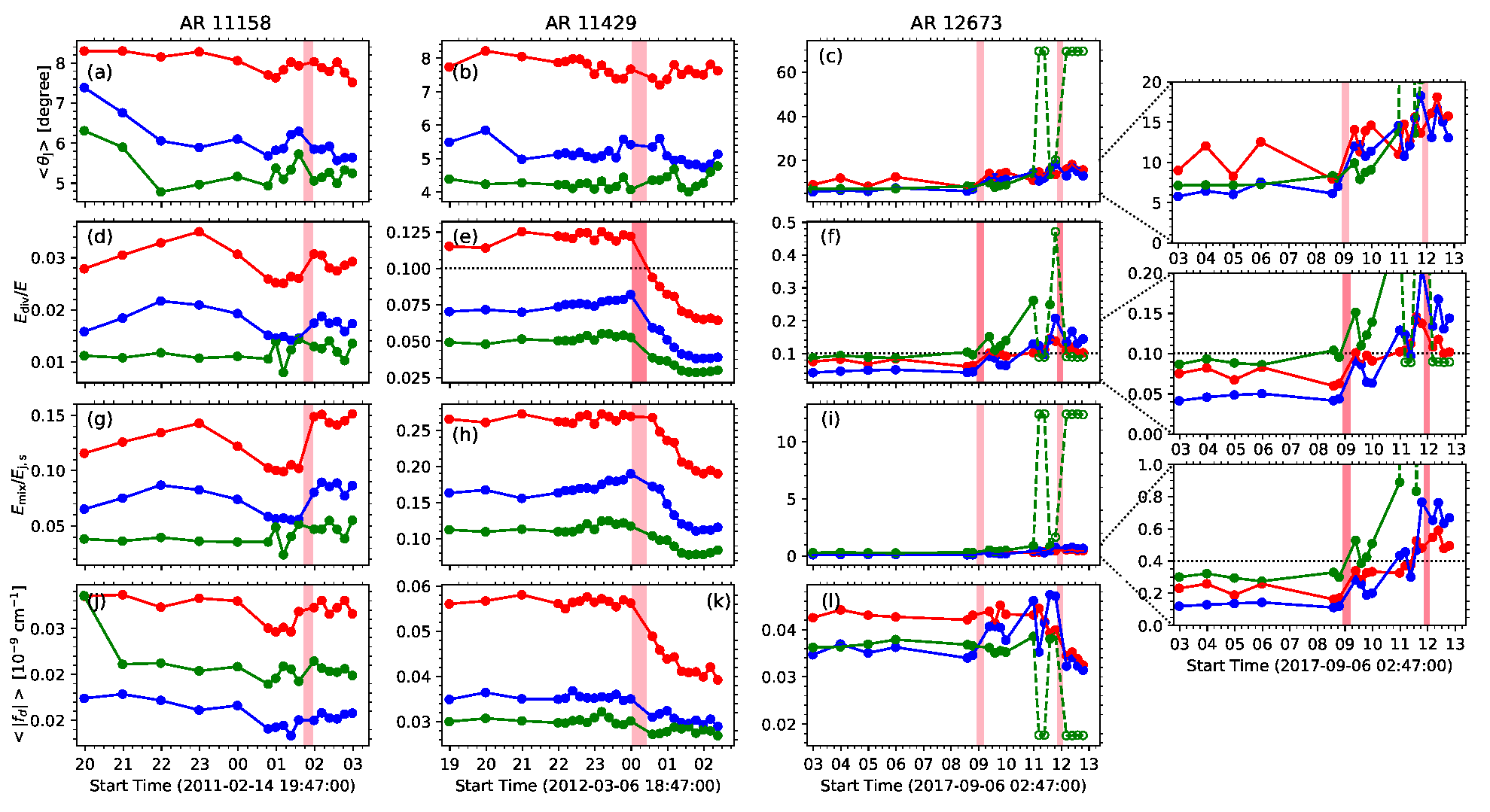}
\caption{Quality of the NLFF solutions. From top to bottom for the individual ARs, we show: the volume-averaged current-weighted angle, $\tjavg$, the ratio of nonsolenoidal-to-total magnetic energy, $\Edivprime$, the ratio $\Emixprime$, and its resolution-invariant complement, $\fdavg$. Red, blue, and green colors correspond to bin1, bin2, and bin4 data, respectively, with inherent plate scales of $\sim$0.36, $\sim$0.72, and $\sim$1.4~Mm. Empty symbols mark disqualifying NLFF solutions (see Sect.~\ref{ss:nlff_quality} for explanation). The inlets in the rightmost column show a subrange of values for enhanced visibility. Vertical bars indicate the impulsive phase of X-class flares that occurred during the analyzed time intervals.}
\label{fig:fig2}
\end{figure*}
%%%%%%%%%%%%%%%%%%%%%%%%%%%%%%%%%%%%%%%%%%%%%%%%%%%%%%%%%%%%%%%%%%%%%%%%%%%%%%%%%%%%%

The NLFF %\so{time-series} 
{time series} of ARs 11158 and 11429 (Fig.~\ref{fig:fig2}(a) and (b), respectively) exhibit values of $\tjavg$$\lesssim$$[10\degree,7\degree,6\degree]$ for the bin[1,2,4]-based solutions, implying that the force-free quality is systematically higher %\so{at lower spatial resolution} 
{for larger pixel sizes}. The overall changes computed for the bin[2,4]-based %\so{time-series} 
{time series} are $\delta\tjavg$$\simeq$[$-$24.1$\pm$4.6,$-$33.4$\pm$4.7]\% for AR~11158 and $\simeq$[$-$32.8$\pm$4.2,$-$44.0$\pm$3.2]\% for AR~11429. Though not shown explicitly, we note that similar findings are obtained from the analysis of $\cwavg$. The situation is different for AR~12673, for which no such systematic improvement of $\tjavg$ %\so{at successively lower spatial resolutions} 
{for a successively larger pixel size} is observed (Fig.~\ref{fig:fig2}(c)). While the bin2-based solutions exhibit comparably lower values of $\tjavg$ before the occurrence of the first X-class flare (before $\sim$09:00~UT), it is the bin4-based solutions that do so after its occurrence (between $\sim$09:24 and 10:00~UT; see inlet to Fig.~\ref{fig:fig2}(c)).

From the decomposition of the magnetic energy,  for AR~11158 we find values of $\Edivprime$$\lesssim$[0.04,0.03,0.02] for the bin[1,2,4]-based NLFF models, respectively (Fig.~\ref{fig:fig2}(d)), that is, the solutions %\so{at lower spatial resolution} 
{based on data at larger pixel sizes} are of higher solenoidal quality. Similarly, we find values of $\Edivprime$$\lesssim$[0.13,0.09,0.06] for the bin[1,2,4]-based NLFF models for AR~11429 (Fig.~\ref{fig:fig2}(e)). More precisely, the overall changes across the bin[2,4]-based %\so{time-series} 
{time series} with respect to the corresponding bin1-based %\so{time-series} 
{time series} are $\delta\Edivprime$$\simeq$[$-$40.6$\pm$3.4,$-$58.3$\pm$7.5]\% for AR~11158 and $\simeq$[$-$38.8$\pm$3.0,$-$57.2$\pm$1.5]\% for AR~11429, which means we observe a comparable overall improvement of solenoidal quality %\so{at}
{for} successively %\so{lower spatial resolution} 
{larger pixel sizes} for the two ARs. This is different for AR~12673 where the bin2-based series appears to be that of highest solenoidal quality, followed by the bin1- and bin4-based series (Fig.~\ref{fig:fig2}(f)), suggesting that $\Edivprime$ does not scale with %\so{underlying spatial resolution} 
{plate scale of the underlying model}. 

We also note that $\Ediv=0.1$ is occasionally exceeded within the individual %\so{time-series} 
{time series}, that is, specific solutions may not be suitable for trustworthy subsequent computation of magnetic helicity (see Sect.~\ref{sss:m_quality} for details). For instance, the bin1-based NLFF models of AR~11429  exhibit values of $\Edivprime>0.1$ prior to the occurrence of the X-class flare (see horizontal line in Fig.~\ref{fig:fig2}(e) for reference). For AR~12673, the situation is even more dramatic, where the NLFF models at all three %\so{spatial resolutions} 
{spatial samplings} exhibit values of $\Edivprime>0.1$ at various time instances after the occurrence of the first X-class flare (after $\sim$09:12~UT; see inlet to Fig.~\ref{fig:fig2}(f)).

For the ratio, $\Emixprime$, we find values of $\lesssim$[0.15,0.10,0.06] for the bin[1,2,4]-based NLFF models of AR~11158 (Fig.~\ref{fig:fig2}(g)), and in the range $\lesssim$[0.30,0.20,0.15] for AR~11429 (Fig.~\ref{fig:fig2}(h)), which corresponds to overall changes across the bin[2,4]-based %\so{time-series} 
{time series} of $\delta\Emixprime$$\simeq$[$-$42.2$\pm$3.3,$-$66.2$\pm$7.5]\% and $\simeq$[$-$37.3$\pm$3.9,$-$57.9$\pm$2.3]\%, respectively. As before, for $\Edivprime$, a comparable overall improvement of solenoidal quality at successively %\so{lower spatial resolution} 
{larger pixel sizes} is observed for the two ARs. Comparatively larger values of $\Emixprime$$\lesssim$0.4 are found for AR~12673 prior to the occurrence of the first X-class flare, while increasing to larger values after the occurrence of the second X-class flare (see inlet to Fig.~\ref{fig:fig2}(i)). Notably, extreme values of $\Emixprime$$>$1 are found from the bin4-based solutions. In Sect.~\ref{sss:energies}, we show that those models exhibit a negative free-energy budget (and are therefore dubbed `non-physical') and are observed in conjunction with low force-free quality ($\tjavg$$\gtrsim$20$\degree$; compare Fig.~\ref{fig:fig2}(c)), indicating the failure of successful extrapolation.

The computed values of the fractional flux are in the range $\fdavg$($\times$$10^{9}$\,cm$^{-1}$)$\lesssim$[0.04,0.02,0.03] for the bin[1,2,4]-based NLFF models of AR~11158 (Fig.~\ref{fig:fig2}(j)), $\lesssim$[0.06,0.04,0.03] for AR~11429 (Fig.~\ref{fig:fig2}(k)), and $\lesssim$[0.05,0.05,0.04] for AR~12673 (Fig.~\ref{fig:fig2}(l)). In other words, while for AR~11158 the bin2-based modeling exhibits the lowest values of $\fdavg$, it is the bin4-based modeling that exhibits lowest values  for AR~11429, quite consistently across the corresponding %\so{time-series} 
{time series}. For AR~12673, again, NLFF models that are qualitatively superior at distinct time instances are not necessarily associated to a systematically different %\so{spatial resolution} 
{plate scale}. In fact, higher and lower values of $\fdavg$ tend to be found together with larger and smaller values of $\tjavg$, respectively (compare Fig.~\ref{fig:fig2}(c)). The problematic bin4-based solutions of AR~12673 (for which extreme values of $\tjavg$ and $\Emixprime$ and partly $\Edivprime$ were also found) exhibit the lowest values ($\fdavg\lesssim0.02$; see symbols marked by empty circles in Fig.~\ref{fig:fig2}(l)), which, on the contrary, suggest a high solenoidal quality.

%%%%%%%%%%%%%%%%%%%%%%%%%%%%%%%%%%%%%%%%%%%%%%%%%%%%%%%%%%%%%%%%%%%%%%%%%%%%%%%%%%%%%
\begin{figure*}
\centering
\includegraphics[width=0.9\textwidth]{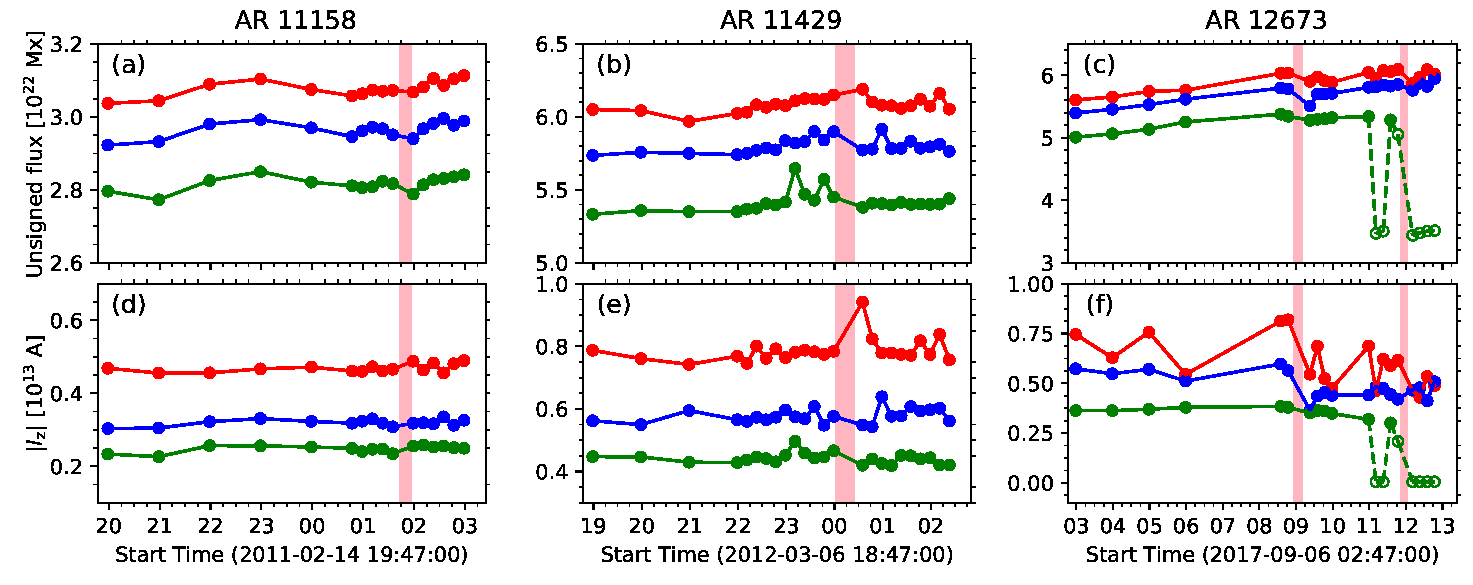}
\caption{Area-integrated parameters computed from the NLFF lower boundary data {with different inherent pixel sizes} as a function of time for three ARs. %\so{, and at three different spatial resolutions.} 
{\it Top:} Unsigned magnetic flux, $\funsign$. {\it Bottom:} Unsigned vertical current, $\izunsign$. Red, blue, and green colors correspond to bin1, bin2, and bin4 data, respectively, with inherent plate scales of $\sim$0.36, $\sim$0.72, and $\sim$1.4~Mm. Vertical bars indicate the impulsive phase of X-class flares that occurred during the time intervals analyzed.}
\label{fig:fig3}
\end{figure*}
%%%%%%%%%%%%%%%%%%%%%%%%%%%%%%%%%%%%%%%%%%%%%%%%%%%%%%%%%%%%%%%%%%%%%%%%%%%%%%%%%%%%%

%%%%%%%%%%%%%%%%%%%%%%%%%%%%%%%%%%%%%%%%%%%%%%%%%%%%%%%%%%%%%%%%%%%%%%%%%%%%%%%%%%%%
\subsection{NLFF-model-deduced quantities}\label{ss:nlff_phys}
%%%%%%%%%%%%%%%%%%%%%%%%%%%%%%%%%%%%%%%%%%%%%%%%%%%%%%%%%%%%%%%%%%%%%%%%%%%%%%%%%%%%

In the following, we analyze physical quantities deduced from the %\so{NLFF model time-series} 
{time-series NLFF modeling} %\so{at different spatial resolutions} 
{with three different plate scales}. As above, we do so by evaluating time-series-averaged changes computed from Eq.~(\ref{eq:tmean}), albeit disregarding nonphysical solutions ($\Efree<0$) within the individual %\so{time-series} 
{time series} (marked by empty plot symbols in Figs.~\ref{fig:fig3} and \ref{fig:fig4}). In addition, when analyzing the magnetic helicity in Sect.~\ref{sss:helicities} we also disregard NLFF solutions with non-negligible solenoidal errors ($\Edivprime>0.1$; marked by empty plot symbols in Figs.~\ref{fig:fig5}).

%%%%%%%%%%%%%%%%%%%%%%%%%%%%%%%%%%%%%%%%%%%%%%%%%%%%%%%%%%%%%%%%%%%%%%%%%%%%%%%%%%%%%
\subsubsection{Unsigned flux and current}\label{sss:fluxes_currents}
%%%%%%%%%%%%%%%%%%%%%%%%%%%%%%%%%%%%%%%%%%%%%%%%%%%%%%%%%%%%%%%%%%%%%%%%%%%%%%%%%%%%%

From the lower boundaries of the bin[2,4]-based NLFF model, using the bin1-based %\so{time-series} 
{time series} as the base against which to compute the percentage changes, for AR~11158 we find $\delta\funsign$$\simeq$[$-$3.6$\pm$0.3,$-$8.5$\pm$0.4]\% (Fig.~\ref{fig:fig3}(a)). For AR~11429, the corresponding changes are $\delta\funsign$$\simeq$[$-$4.7$\pm$0.8,$-$11.0$\pm$1.1]\% (Fig.~\ref{fig:fig3}(b)) and for AR~12673 they are $\delta\funsign$$\simeq$[$-$3.5$\pm$1.2,$-$10.8$\pm$1.0]\% (Fig.~\ref{fig:fig3}(c)). Furthermore,  we find $\delta\izunsign$$\simeq$[$-$31.8$\pm$2.6,$-$47.1$\pm$2.1]\% for AR~11158 (Fig.~\ref{fig:fig3}(d)), $\simeq$[$-$26.4$\pm$4.7,$-$43.7$\pm$3.6]\% for AR~11429 (Fig.~\ref{fig:fig3}(e)), and $\simeq$[$-$17.6$\pm$14.8,$-$43.7$\pm$10.1]\% for AR~12673 (Fig.~\ref{fig:fig3}(f)).

We note that the lower boundary data for the final NLFF model  necessarily differ from the input data (cf.\ Sect.~\ref{ss:data}) because they are altered once during the preprocessing step and also iteratively updated during the optimization process (for details see Sect.~\ref{ss:nlff}). As a consequence, $\funsign$ computed from the lower boundaries of the NLFF model  is lower than in the corresponding input data (up to $\approx$10\% at most), as is $\izunsign$ (up to $\approx$40\% at most). Changes to the input data, especially to the horizontal magnetic field components, are expected given their inconsistency with the force-free assumption; also, they are known to be substantial in comparison to the uncertainties of the input data \citep[for a corresponding analysis, see e.g., Sect.~3.2 of][]{2015ApJ...811..107D}. However, qualitatively, the trends within the individual NLFF lower boundary-based %\so{time-series} 
{time series} are similar to those in the corresponding {time series of} input data %\so{time-series} 
(with the exception of those that stem from unphysical NLFF solutions) and the induced changes are on the order of the %\so{resolution-induced} 
{spatial-sampling-induced} changes as listed earlier.

%%%%%%%%%%%%%%%%%%%%%%%%%%%%%%%%%%%%%%%%%%%%%%%%%%%%%%%%%%%%%%%%%%%%%%%%%%%%%%%%%%%%%
\subsubsection{Magnetic energy}\label{sss:energies}
%%%%%%%%%%%%%%%%%%%%%%%%%%%%%%%%%%%%%%%%%%%%%%%%%%%%%%%%%%%%%%%%%%%%%%%%%%%%%%%%%%%%%

%%%%%%%%%%%%%%%%%%%%%%%%%%%%%%%%%%%%%%%%%%%%%%%%%%%%%%%%%%%%%%%%%%%%%%%%%%%%%%%%%%%%%
\begin{figure*}
\centering
\includegraphics[width=0.9\textwidth]{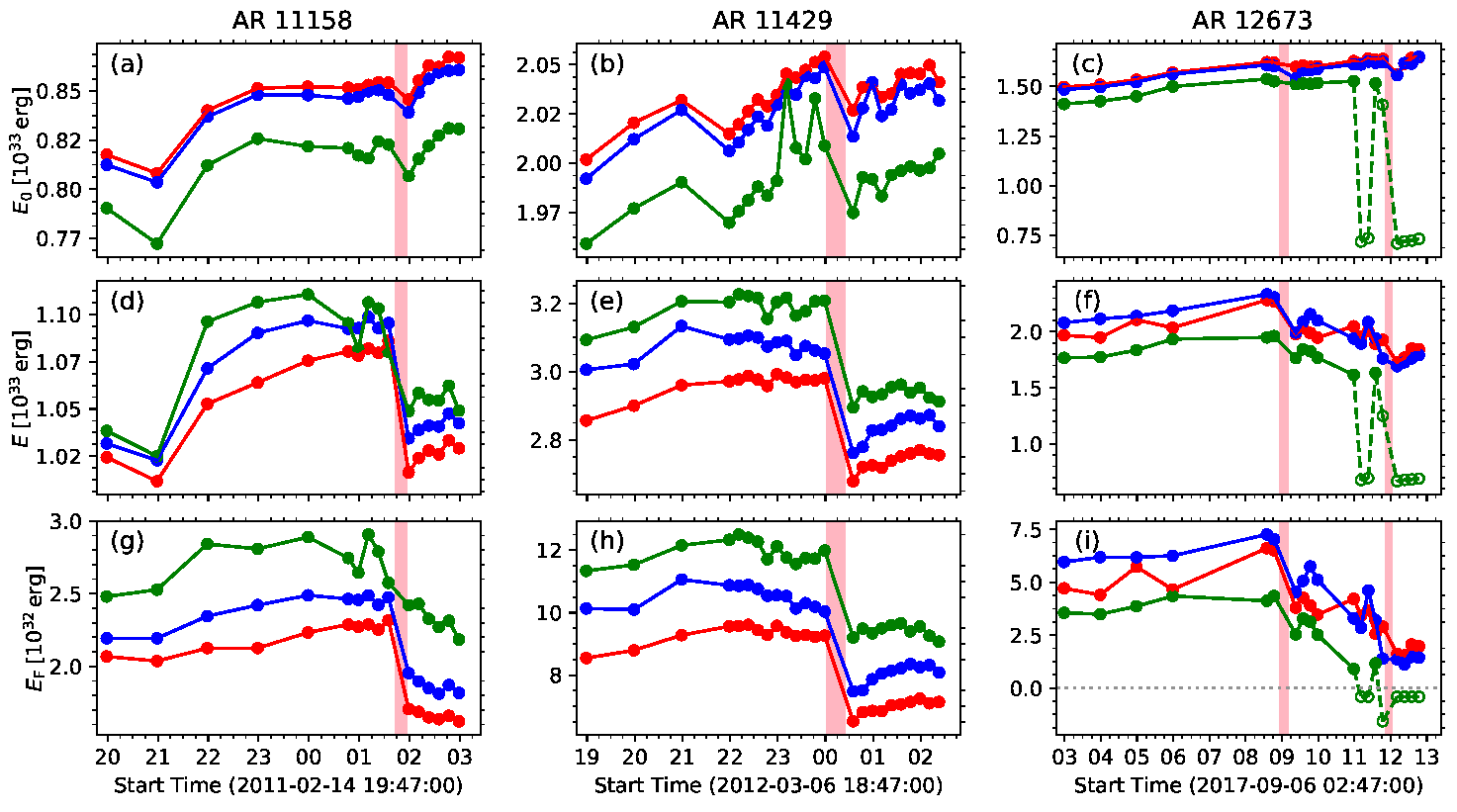}
\caption{Volume-integrated magnetic energies computed from the NLFF solutions {with three different plate scales} as a function of time for three ARs. %\so {, and at three different spatial resolutions.}
{\it Top:} Potential energy, $\Epot$. {\it Middle:} Total energy, $\Etot$. {\it Bottom:} Free magnetic energy, $\Efree=\Etot-\Epot$. Red, blue, and green colors correspond to bin1, bin2, and bin4 data, respectively, with inherent plate scales of $\sim$0.36, $\sim$0.72, and $\sim$1.4~Mm. Vertical bars indicate the impulsive phase of X-class flares that occurred during the analyzed time intervals.}
\label{fig:fig4}
\end{figure*}
%%%%%%%%%%%%%%%%%%%%%%%%%%%%%%%%%%%%%%%%%%%%%%%%%%%%%%%%%%%%%%%%%%%%%%%%%%%%%%%%%%%%%

For all three considered ARs, the computed potential field energies, $\Epot$, are smaller %\so{at reduced spatial resolution} 
{for larger pixel sizes} (Fig.~\ref{fig:fig4}(a)--(c)). More quantitatively, in comparison to $\Epot$ from the corresponding bin1-based NLFF model %\so{time-series} 
{time series}, the changes that are found are
%\LEt{Please check that I have retained your intended meaning.} 
$\delta\Epot$$\simeq$[$-$0.6$\pm$0.2,$-$4.0$\pm$0.5]\% for AR~11158, $\simeq$[$-$0.4$\pm$0.1,$-$2.1$\pm$0.5]\% for AR~11429, and $\simeq$[$-$0.9$\pm$0.8,$-$5.6$\pm$0.7]\% for AR~12673, for the bin[2,4]-based %\so{time-series} 
{time series}, respectively.

Overall larger values of total magnetic energy, $\Etot$, are found %\so{at lower spatial resolution} 
{for larger pixel sizes} for ARs 11158 and 11429 (Fig.~\ref{fig:fig4}(d) and (e), respectively). More quantitatively, in comparison to the bin1-data based NLFF models, we find changes of $\delta\Etot$$\simeq$[1.4$\pm$0.4,2.3$\pm$1.3]\% for AR~11158, and $\simeq$[3.7$\pm$0.8,7.4$\pm$0.8]\% for AR~11429 at bin[2,4]. In contrast, and similar to all other analyzed quantities so far, no such systematic dependence of $\Etot$ on underlying %\so{spatial resolution} 
{spatial sampling} is found for AR~12673 (Fig.~\ref{fig:fig4}(f)). Here, we find $\delta\Etot$$\simeq$1.3$\pm$5.0\% for the bin2-based and $\delta\Etot$$\simeq$$-$11.5$\pm$4.1\% for the bin4-based NLFF model %\so{time-series} 
{time series}. 

For the free magnetic energies, $\Efree$, we note trends similar to that found for the corresponding values of $\Etot$  (see Fig.~\ref{fig:fig4}(g)--(i)). More quantitatively, overall and in comparison to the bin1-data based NLFF models, the bin[2,4]-based estimates are found to be $\delta\Efree$$\simeq$[10.2$\pm$2.7,29.8$\pm$10.0]\% for AR~11158 and $\simeq$[13.9$\pm$2.9,31.2$\pm$4.7]\% for AR~11429. In contrast, we find $\simeq$[6.3$\pm$29.4,$-$32.7$\pm$18.6]\% for AR~12673 (Fig.~\ref{fig:fig4}(i)). The nonphysical solutions, such as the bin4-based NLFF models of AR~12673 at the end of the considered time interval where $\Efree<0$ (see empty plot symbols in Fig.~\ref{fig:fig4}(i)), were previously identified based on outstandingly poor NLFF model quality metrics (see Sect.~\ref{ss:nlff_quality} for details) and were not considered for computation of the percentages above. 

We note that the %\so{spatial-resolution-induced} 
{spatial-sampling-induced} changes to $\Epot$ reflect those observed for $\funsign$. This may be expected as the potential field is determined from the vertical field on the lower boundary of the model volume. In contrast, $\Etot$ and $\Efree$ show a different behavior from the supposedly indicative unsigned vertical current ($\izunsign$). Naively, one might expect to find larger corresponding values for larger values of $\izunsign$, as the latter represents a measure of enhanced complexity in the horizontal field. Though it is true that %\so{a higher spatial resolution} 
{smaller pixel sizes on} overall %\so{relates} 
{relate} to higher values of $\izunsign$ (cf.\ Fig.~\ref{fig:fig3}), this is not true for $\Etot$. Consequently, this is also not true for the free magnetic energy (compare Fig.~\ref{fig:fig4}(d)--(i)) as it is calculated as $\Efree=\Etot-\Epot$, and $\Epot$ is larger %\so{at higher spatial resolutions} 
{for smaller pixel sizes}. For completeness, we note here that we also inspected the volume-integrated total unsigned current, $\Itotunsign$, in order to better understand the obtained total energies. Though not shown explicitly, trends throughout the individual %\so{time-series} 
{time series} as well as %\so{resolution-induced} 
{spatial-sampling-induced} changes are found to be very similar to that of $\Etot$ and $\Efree$, namely larger integrated values %\so{at lower spatial resolutions} 
{for larger pixel sizes}.

This apparent discrepancy can partly be resolved by comparison to the solenoidal quality of the NLFF models in Fig.~\ref{fig:fig2}, revealing a rather clear dependency. Both trends within the individual %\so{time-series} 
{time series} (at different %\so{spatial resolutions} 
{plate scales}) and %\so{resolution-induced} 
changes %\so{for time-series of individual ARs} 
{to them} are found to be {directly reflected}
in the %\so{time-series} 
{time series} of $\Emixprime$ (and to a somewhat lesser extent in the %\so{time-series} 
{time series} of $\Edivprime$ in conjunction with $\tjavg$). More precisely, volume-integrated energies (and unsigned currents) are higher for NLFF models of higher solenoidal quality. For instance, the bin4-based NLFF models of ARs~11158 and 11429 exhibit larger values of $\Etot$ and $\Efree$ (and $\Itotunsign$) in conjunction with lowest values of $\Emixprime$ than at other %\so{spatial resolutions} 
{plate scales}. In contrast, the bin4-based NLFF models of AR~12673 exhibit lower energies than %\so{at other spatial resolutions} 
{for other pixel sizes} (bin1 and bin2) and simultaneously exhibit larger values of $\Emixprime$.

%%%%%%%%%%%%%%%%%%%%%%%%%%%%%%%%%%%%%%%%%%%%%%%%%%%%%%%%%%%%%%%%%%%%%%%%%%%%%%%%%%%%%
\begin{figure*}
\centering
\includegraphics[width=0.9\textwidth]{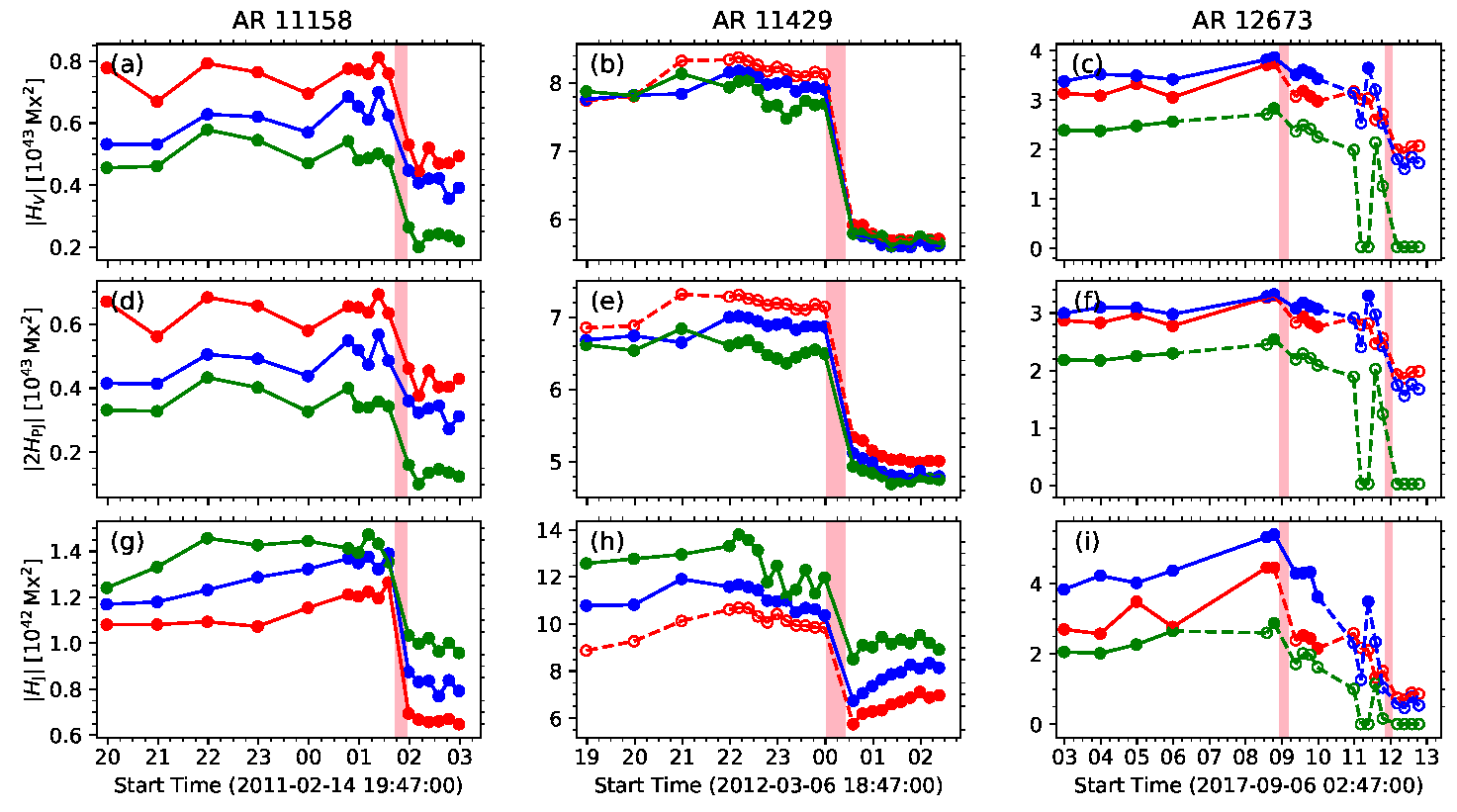}
\caption{Volume-integrated relative helicities computed from the NLFF solutions {with three different plate scales} as a function of time for three ARs. %\so{, and at three different spatial resolutions.}
{\it Top:} Total helicity, $\Hv$. {\it Middle:} Volume-threading helicity, $\Hpj$. {\it Bottom:} Helicity of the current-carrying field, $\Hj$. Red, blue, and green color correspond to bin1, bin2, and bin4 data, respectively, with inherent plate scales of $\sim$0.36, $\sim$0.72, and $\sim$1.4~Mm. Vertical bars indicate the impulsive phase of X-class flares that occurred during the analyzed time intervals.}
\label{fig:fig5}
\end{figure*}
%%%%%%%%%%%%%%%%%%%%%%%%%%%%%%%%%%%%%%%%%%%%%%%%%%%%%%%%%%%%%%%%%%%%%%%%%%%%%%%%%%%%%

%%%%%%%%%%%%%%%%%%%%%%%%%%%%%%%%%%%%%%%%%%%%%%%%%%%%%%%%%%%%%%%%%%%%%%%%%%%%%%%%%%%%%
\subsubsection{Magnetic helicity}\label{sss:helicities}
%%%%%%%%%%%%%%%%%%%%%%%%%%%%%%%%%%%%%%%%%%%%%%%%%%%%%%%%%%%%%%%%%%%%%%%%%%%%%%%%%%%%%

For the total relative helicity, $\Hv$, overall and in comparison to the bin1-data-based NLFF models the bin[2,4]-based estimate changes are found to be $\delta\Hv$$\simeq$[$-$17.9$\pm$5.6,$-$40.7$\pm$9.9]\% for AR~11158 and $\simeq$[$-$1.6$\pm$0.7,$-$0.7$\pm$1.0]\% for AR~11429 (Fig.~\ref{fig:fig5}(a) and (d), respectively). Similarly, one finds $\delta\Hpj$$\simeq$[$-$23.7$\pm$6.3,$-$52.4$\pm$12.7]\% for AR~11158 and $\simeq$[$-$4.1$\pm$0.8,$-$5.9$\pm$1.2]\% for AR~11429, respectively (Fig.~\ref{fig:fig5}(b) and (e), respectively). In contrast, comparatively lower values of the current-carrying helicities, $\Hj$, are found %\so{at higher spatial resolutions} 
{for smaller pixel sizes} (Fig.~\ref{fig:fig5}(g) and (h), respectively), with $\delta\Hj$$\simeq$[16.5$\pm$6.7,31.7$\pm$15.8]\% and $\simeq$[17.9$\pm$2.6,39.5$\pm$7.2]\%, respectively. In contrast, and similar to all other quantities analyzed for AR~12673 so far, no apparent dependencies of the relative helicities on the underlying %\so{spatial resolution} 
{plate scale} are found. Instead, $\delta\Hv$$\simeq$[9.8$\pm$5.2,$-$22.6$\pm$3.8]\% (Fig.~\ref{fig:fig5}(c)), $\delta\Hpj$$\simeq$[6.2$\pm$4.0,$-$22.4$\pm$2.9]\% (Fig.~\ref{fig:fig5}(f)), and $\delta\Hj$$\simeq$[48.5$\pm$24.4,$-$24.0$\pm$12.8]\% (Fig.~\ref{fig:fig5}(i)) for the bin[2,4]-based modeling, respectively. 

Again, comparison to the solenoidal quality of the underlying NLFF models (Fig.~\ref{fig:fig2}) reveals a dependency of the relative helicities on %\so{spatial resolution} {the underlying pixel size},
%\LEt{Please check that I have retained your intended meaning.} 
very similar to that found for the magnetic energies (Sect.~\ref{sss:energies}). This can be seen from the bin4-based NLFF models of AR~12673, which yield lower values for the relative helicities and simultaneously larger values of $\Emixprime$ than the corresponding models at %\so{higher spatial resolutions} {lower plate scales} (bin2 and bin1). 

%%%%%%%%%%%%%%%%%%%%%%%%%%%%%%%%%%%%%%%%%%%%%%%%%%%%%%%%%%%%%%%%%%%%%%%%%%%%%%%%%%%%%
%%%%%%%%%%%%%%%%%%%%%%%%%%%%%%%%%%%%%%%%%%%%%%%%%%%%%%%%%%%%%%%%%%%%%%%%%%%%%%%%%%%%%
\section{Discussion} \label{s:discussion}
%%%%%%%%%%%%%%%%%%%%%%%%%%%%%%%%%%%%%%%%%%%%%%%%%%%%%%%%%%%%%%%%%%%%%%%%%%%%%%%%%%%%%
%%%%%%%%%%%%%%%%%%%%%%%%%%%%%%%%%%%%%%%%%%%%%%%%%%%%%%%%%%%%%%%%%%%%%%%%%%%%%%%%%%%%%

We demonstrated that the %\so{resolution-induced} 
{spatial-sampling-induced} effects are not only different at different times (for a specific AR) but are also distinctly different for different ARs. The chosen setup in this study (i.e., we employ three {time series of} NLFF %\so{time-series} 
{models} %\so{at different spatial resolutions} 
{with different plate scales} for three different ARs) furthermore allows us to study overall trends to be expected for NLFF modeling (and subsequent magnetic energy and helicity computations). Therefore, we %\so{calculate}
{generate} histograms of the changes to the physical %\so{variables} 
{quantities} %\so{due to a reduction of the spatial resolution} 
{caused by the down-sampling of the input data.} %\so{, that is,} 
{To do so,} we %\so{compute} 
{quantify} the %\so{differences} 
{variations of the physical quantities} %\so{between} 
{by comparison of} all qualifying bin2- and bin4-based NLFF solutions. %\so{, if a corresponding bin1-based NLFF model qualifies for comparison (i.e., 
%\LEt{ the meaning here is unclear; please consider alternative wording.} suffices the same quality criteria). 
{We only do so if all of the three solutions at a given time instant meet our quality criteria.}  Generally, a NLFF model qualifies if it is physical ({i.e., if} $\Efree$$>$0; 59 NLFF models at bin2 and 52 at bin4). It qualifies for subsequent helicity-computation if it is sufficiently solenoidal ($\Edivprime$$\leq$0.1; 35 NLFF models at bin2 and 31 at bin4). For those, median values as well as corresponding median absolute deviations for the induced changes (denoted by angular brackets hereafter) are discussed in the following, and are interpreted in context with the changes to the individual NLFF model %\so{time-series}
 {time series} listed in Sect.~\ref{s:results}.

\begin{figure*}
\centering
\includegraphics[width=0.9\textwidth]{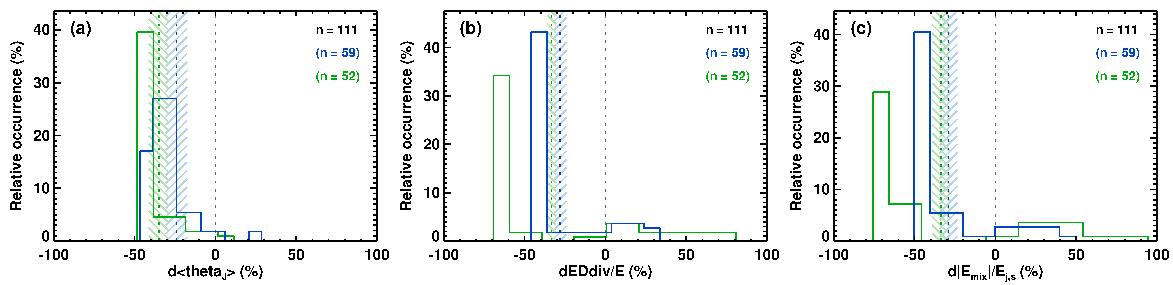}
\caption{Histograms displaying the variations of the NLFF model quality as quantified by (a) $\tjavg$, (b) $\Edivprime$, and (c) $\Emixprime$. Relative differences between the bin1-based estimates and the bin2-(blue) and bin4-based (green) modeling  are shown. The total number of considered (qualifying) NLFF models is indicated in black and is used as the basis to compute percentages. The total number of qualifying solutions at bin2 and bin4 are indicated in blue and green, respectively. Median values and median absolute deviation derived from the histograms are shown as dashed vertical lines and shaded bars, respectively.}
\label{fig:fig6}
\end{figure*}

A first main finding of our analysis regards the relative power of distinct metrics to measure the quality of NLFF models. When using the most indicative (sensitive) metrics for a corresponding quantification, we find that, overall, the NLFF model quality is higher %\so{at reduced spatial resolutions} 
{for larger pixel sizes}. Using $\tjavg$ as a measure, median changes by $\langle\delta\tjavg\rangle$$\approx$$-$26.4$\pm$3.8\% and $\approx$$-$34.3$\pm$5.2\% are found for the bin2-based and bin4-based modeling, respectively (Fig.~\ref{fig:fig6}(a)), using the bin1-based NLFF modeling as a reference (which is used as a basis for all median changes listed in the following). Though not explicitly shown, we note that the corresponding analysis of $\cwavg$ leads to very similar conclusions. An improvement of the solenoidal quality of the NLFF models %\so{at lower spatial resolution} 
{for larger pixel sizes} might be expected, as the application of binning to mimic %\so{the reduction of} 
{a lower} spatial resolution reduces gradients present in the original data, that is, it should yield a reduction of $\nabla\cdot\vB$. Corresponding conclusions can be drawn from the quantities most sensitive to the solenoidal quality of a magnetic field, $\Edivprime$ and $\Emixprime$. Here, median changes are found of $\langle\delta\Edivprime\rangle$$\approx$[$-$37.8$\pm$3.7,$-$43.7$\pm$6.1]\%  (Fig.~\ref{fig:fig6}(b)) and $\langle\delta\Emixprime\rangle$$\approx$[$-$39.7$\pm$3.0,$-$45.9$\pm$7.2]\% (Fig.~\ref{fig:fig6}(c)) for the bin[2,4]-based modeling, respectively.
%\LEt{Please check that I have retained your intended meaning.}. 
Those two measures were found to be most sensitive (and indicative) regarding the solenoidal quality of the tested NLFF models in Sect.~\ref{ss:nlff_quality} \citep[see also][]{2019ApJ...880L...6T}, and to be superior to the use of an alternative metrics to quantify the divergence-freeness, such as $\fiavg$ and $\fdavg$. 

\cite{2015ApJ...811..107D} used $\fiavg$ as a measure to quantify the divergence-free quality at different spatial resolutions, according to which the analyzed optimization-based modeling exhibited successively larger values at lower spatial resolutions (see their Table~2). If we were to draw conclusions based on $\fiavg$, we would base these on the corresponding median values of $\langle\fiavg\times10^4\rangle$$\approx$[2.5$\pm$0.5,3.7$\pm$0.4,6.9$\pm$0.5] for bin[1,2,4]-based modeling, respectively, and we would arrive at a similar conclusion to that of \cite{2015ApJ...811..107D}, namely that the divergence-free property is improved at increased spatial resolution. Nevertheless, $\fiavg$ has recently been dubbed inappropriate for the purpose of analyzing resolution-induced aspects by \cite{2020ApJ...900..136G}, who proposed an improved (refined) corresponding measure, $\fdavg$, which is almost insensitive to the %\so{spatial resolution} 
{plate scale} of the analyzed NLFF solution. Here, we find median values of $\langle\fdavg\times10^{11}{\rm cm}^{-1}\rangle$$\approx$[4.2$\pm$0.9,3.1$\pm$0.3,3.0$\pm$0.3] for bin[1,2,4]-based modeling, respectively, which translates to the lowest divergence-freeness at highest spatial resolution.
%\LEt{Please check that I have retained your intended meaning.}. 
Moreover, the trend of a lower solenoidal quality %\so{at higher spatial resolution} 
{for smaller pixel sizes} can also be deduced from the optimization-based NLFF models studied by \cite{2015ApJ...811..107D} (see their Table~4), and we can obtain average estimates for plate scales that approximately correspond to those used in our study ([0.36,0.72,1.44]~Mm for our bin[1,2,4] cases, respectively). In particular, we compute $\langle\left[(\Edivprime)_{\rm bin6},(\Edivprime)_{\rm bin8}\right]\rangle$$\simeq$0.08 (corresponding to an average plate scale of $\sim$0.74~Mm) and use the $(\Edivprime)_{\rm bin14}$$=$0.06 provided by these latter authors (corresponding to an average plate scale of $\sim$1.48~Mm) to find a solenoidal quality improved by $\approx$12\% and $\approx$33\%, respectively, with respect to their $\langle\left[(\Edivprime)_{\rm bin3},(\Edivprime)_{\rm bin4}\right]\rangle$$\simeq$0.09 (corresponding to an average plate scale of $\sim$0.37~Mm). Doing the same for $\Emixprime$, that is, using $(\Emixprime)_{\rm bin14}$$\simeq$1.00 and $\langle\left[(\Emixprime)_{\rm bin6},(\Emixprime)_{\rm bin8}\right]\rangle$$\simeq$1.17, we find improvements of $\approx$3\% and $\approx$17\%, respectively, with respect to their $\langle\left[(\Emixprime)_{\rm bin3},(\Emixprime)_{\rm bin4}\right]\rangle$$\simeq$1.20.

%%%%%%%%%%%%%%%%%%%%%%%%%%%%%%%%%%%%%%%%%%%%%%%%%%%%%%%%%%%%%%%%%%%%%%%%%%%%%%%%%%%%%
%%%%%%%%%%%%%%%%%%%%%%%%%%%%%%%%%%%%%%%%%%%%%%%%%%%%%%%%%%%%%%%%%%%%%%%%%%%%%%%%%%%%%

%%%%%%%%%%%%%%%%%%%%%%%%%%%%%%%%%%%%%%%%%%%%%%%%%%%%%%%%%%%%%%%%%%%%%%%%%%%%%%%%%%%%%
\begin{figure*}
\centering
\includegraphics[width=0.9\textwidth]{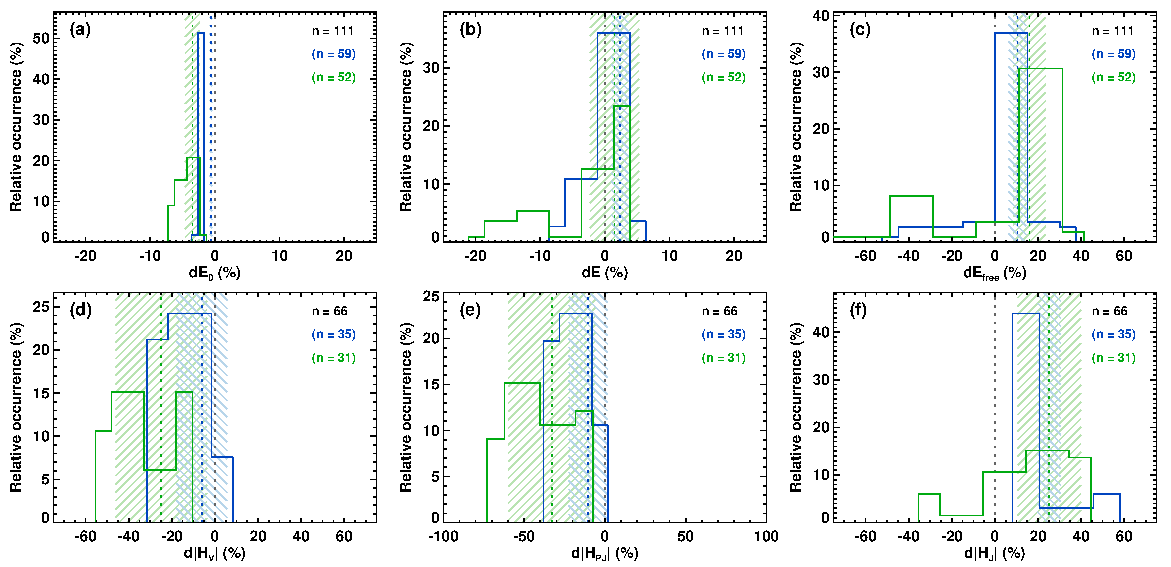}
\caption{Histograms displaying the variations of the volume-integrated magnetic energies $\Epot$ (a), $\Etot$ (b), and $\Efree$ (c) and absolute relative helicities, $\Hv$ (d), $\Hpj$ (e), and $\Hj$ (f). Relative differences between the bin1-based
estimates  and the bin2- (blue) and bin4-based (green) modeling are shown. The total number of considered (qualifying) NLFF models is indicated in black and used as the basis to compute percentages. The total number of qualifying solutions at bin2 and bin4 are indicated in blue and green, respectively. Median values and median absolute deviation derived from the histograms are shown as dashed vertical lines and shaded bars, respectively.}
\label{fig:fig7}
\end{figure*}
%%%%%%%%%%%%%%%%%%%%%%%%%%%%%%%%%%%%%%%%%%%%%%%%%%%%%%%%%%%%%%%%%%%%%%%%%%%%%%%%%%%%%
 
As a second major finding, we may state that there are certain overall tendencies regarding how a change in the %\so{spatial resolution} 
{spatial sampling} translates into a corresponding variation in the deduced physical quantities. Overall, the NLFF lower boundary area-integrated quantities, $\funsign$ and $\izunsign$, exhibit %\so{resolution-dependent}
{spatial-sampling-dependent} variations in the form of a successive reduction when the %\so{resolution is reduced} 
{pixels size is enlarged} (see Sect.~\ref{sss:fluxes_currents} and Fig.~\ref{fig:fig3}). This is actually expected because the binning, which is used to mimic a reduction %\so{in} 
{of} the spatial resolution of the data, necessarily reduces amplitudes and gradients with respect to that of the original-resolution data. Though not shown explicitly, we deduce median changes of $\langle\delta\funsign\rangle$$\approx$[$-$3.9$\pm$0.4,$-$9.6$\pm$0.9]\% and $\langle\delta\izunsign\rangle$$\approx$[$-$27.3$\pm$4.7,$-$46.4$\pm$2.4]\% 
for the bin[2,4]-based lower boundary data, respectively. The comparatively larger modifications to $\izunsign$ (compared to that of $\funsign$) are expected, because during NLFF modeling the horizontal magnetic field components are altered to a much greater degree than the vertical magnetic field component (hence is $\funsign$; see Sect.~\ref{sss:fluxes_currents} for details). 

Intuitively, corresponding to the lower values of $\funsign$ and $\izunsign$, we would expect to also find smaller values for the volume-integrated estimates (magnetic energies, electric currents, and magnetic helicities) at successively lower spatial resolution. Nevertheless, this is only partially true. For instance, the %\so{resolution-induced} 
{spatial-sampling-induced} changes to $\Epot$ are consistent with those found for $\funsign$, namely $\langle\delta\Epot\rangle$$\approx$[$-$0.6$\pm$0.2,$-$3.7$\pm$1.0]\% for bin[2,4]-based NLFF modeling, respectively (Fig.~\ref{fig:fig7}(a)). This is also true for the total helicity, with $\langle\delta|\Hv|\rangle$$\approx$[$-$5.9$\pm$13.2,$-$25.6$\pm$14.4]\% (Fig.~\ref{fig:fig7}(d)), and the volume-threading helicity with $\langle\delta\Hpj\rangle$$\approx$[$-$10.3$\pm$13.1,$-$33.5$\pm$19.4]\% (Fig.~\ref{fig:fig7}(e)), and is also consistent with the corresponding trends seen in $\langle\delta\izunsign\rangle$. In contrast, weaker electric currents do not necessarily translate to systematically lower volume-integrated total energies (Fig.~\ref{fig:fig7}(b)), free magnetic energies (Fig.~\ref{fig:fig7}(c)), and current-carrying helicities (Fig.~\ref{fig:fig7}(f)), for which successively larger median values are found, namely $\langle\delta\Etot\rangle$$\approx$[3.0$\pm$1.0,1.6$\pm$3.3]\%, $\langle\delta\Efree\rangle$$\approx$[16.0$\pm$4.2,22.0$\pm$8.1]\%, and $\langle\delta\Hj\rangle$$\approx$[25.6$\pm$6.3,24.0$\pm$15.7]\% for bin[2,4]-based modeling, respectively.

 %%%%%%%%%%%%%%%%%%%%%%%%%%%%%%%%%%%%%%%%%%%%%%%%%%%%%%%%%%%%%%%%%%%%%%%%%%%%%%%%%%%%%
\begin{figure*}
\centering
\includegraphics[width=\textwidth]{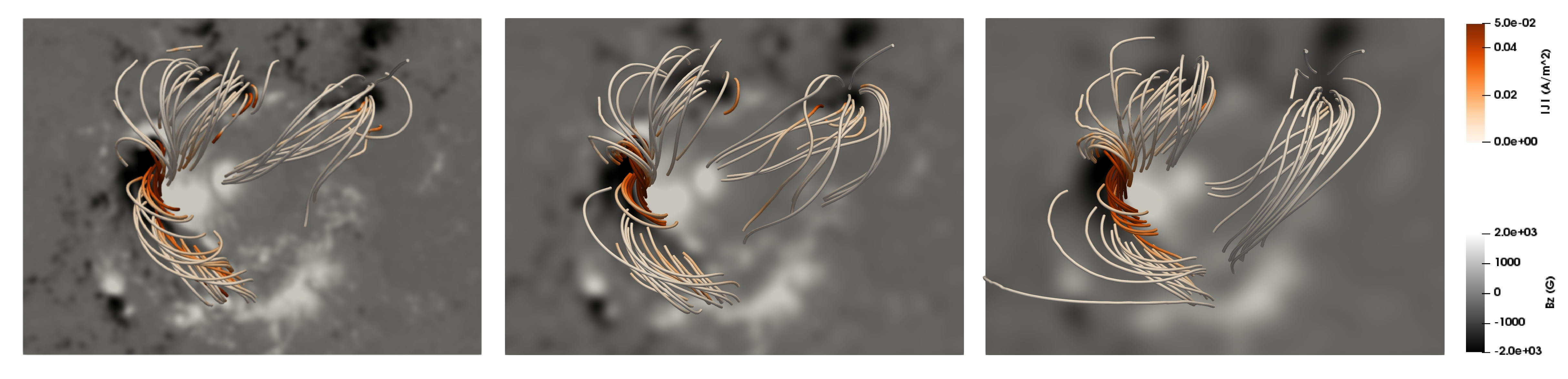}
\caption{Morphology of the reconstructed magnetic field of AR~12673 at 08:48 on 6~September~2017 at different spatial resolutions. From left to right, the bin1-, bin2-, and bin4-based modeling is shown. The same footpoints were used in all cases. Field lines are colored according to the magnitude of the electric current density, $|\vJ|$. The gray-scale background resembles $\Bz$ of the NLFF lower boundary, scaled to $\pm$2~kG.}
\label{fig:fig8}
\end{figure*}
%%%%%%%%%%%%%%%%%%%%%%%%%%%%%%%%%%%%%%%%%%%%%%%%%%%%%%%%%%%%%%%%%%%%%%%%%%%%%%%%%%%%%

\cite{2013ApJ...769...59T}, based on the binning of {\it Hinode}/SOT-SP data to a plate scale of $\sim$0.5~arcsec (called the ``SP$_{\rm bin}$'' case in their study and intended to match the spatial resolution of HMI data in their study) reported a binning-induced decrease in the total unsigned flux and potential field energy, along with an increase in total and free magnetic energy when compared to the NLFF modeling using original-resolution ($\sim$0.3~arcsec at disk center; called ``SP$_{\rm orig}$'' case in their study) SOT-SP data. In contrast, the optimization-based models based on %\so{lower-resolution}
{down-sampled} SOT-SP data in \cite{2015ApJ...811..107D} were associated to lesser total and free magnetic energies. However, as already noted by \cite{2015ApJ...811..107D}, estimates of physical quantities become questionable in the presence of significant residual errors in the divergence of $\vB$. Now, with results from the dedicated studies by {\cite{2016SSRv..201..147V}, \cite{2019ApJ...880L...6T}, and \cite{2020A&A...643A.153T}} at hand, optimization-based NLFF solutions may only be trustworthy if they exhibit values of $\Edivprime$$\lesssim$0.1 and $\Emixprime$$\lesssim$0.4. However, all of the optimization-based models analyzed in the work of \cite{2015ApJ...811..107D} exhibited values of $\Edivprime$$\gtrsim$0.06 and $\Emixprime$$\gtrsim$0.9. Measures of the quality of the analyzed NLFF models were not reported by \cite{2013ApJ...769...59T}, and were therefore not interpreted in context with the obtained estimates of physical parameters. The quality measures for SP-orig model read $\Edivprime$$=$0.08 and $\Emixprime$$=$0.55. For the SP$_{\rm bin}$ model, they read $\Edivprime$$=$0.09 and $\Emixprime$$=$0.51. Thus, observed (apparently %\so{resolution-dependent}
{spatial-sampling-dependent}) trends of deduced model-based physical parameters in \cite{2013ApJ...769...59T} and \cite{2015ApJ...811..107D} must be questioned due to the poor model quality of the underlying NLFF solutions. This makes it difficult to interpret the findings of those studies alongside results from our extended approach and NLFF models of high solenoidal quality $\langle\Edivprime\rangle$$\approx$[0.05$\pm$0.03,0.05$\pm$0.03,0.04$\pm$0.02] and 
$\langle\Emixprime\rangle$$\approx$[0.18$\pm$0.05,0.14$\pm$0.04,0.11$\pm$0.03] for bin[1,2,4]-based modeling. For example, we find that, generally, binning-induced effects include decreases in unsigned magnetic fluxes and potential field energies, along with increases in total and free magnetic energies.

In addition to finding that the solenoidal quality of NLFF solutions directly affects the resulting volume-integrated estimates, the generally lower model quality of NLFF solutions %\so{at higher spatial resolution} 
{with smaller pixel sizes} (Fig.~\ref{fig:fig6}) together with the lower values of volume-integrated quantities directly associated to the presence of electric currents ($\Etot$, $\Efree$ and $\Hj$; Fig.~\ref{fig:fig7}) raises the question of whether one should favor the use of optimization-based NLFF modeling at a reduced spatial resolution. While its use is encouraged for applications to HMI data based on the analysis presented here, a general recommendation in that sense cannot be given, particularly because %\so{resolution-induced} 
{spatial-sampling-induced} effects might behave differently for applications to data from other instruments (e.g., SOT-SP as discussed above). Indeed, a robust understanding of %\so{resolution-induced} 
{spatial-sampling-induced} effects from the application of the optimization method (and of other existing NLFF methods) to SOT-SP data remains elusive.

Another aspect to consider is the following. Trends detected from the %\so{time-series} 
{time series} of qualifying solutions in terms of systematic increases or decreases along individual %\so{time-series} 
{time series} appear consistent across different resolutions, including flare-related changes of magnetic energies and helicities (Figs.~\ref{fig:fig4} and \ref{fig:fig5}). However, the magnitude of those changes decreases with decreasing resolution. For instance, taking the last/first available data point prior to and following the start/end of the nominal flare impulsive phase, respectively, from bin[1,2,4]-based modeling we find  flare-related changes of $\Delta\Efree$$\approx$[25.7,21.4,9.6]\% for the X2.2 flare hosted by AR~11158, $\Delta\Efree$$\approx$[27.9,25.6,21.2]\% for the X5.4 flare hosted by AR~11429, and $\Delta\Efree$$\approx$[38.4,32.9,31.3]\% for the X2.2 flare hosted by AR~12673. (Due to the lack of qualifying NLFF solutions, a corresponding estimate for the X9.3 flare cannot be provided.) Similar tendencies can be deduced for the flare-related changes of $\Hj$, for example. These findings indicate that the binning of data prior to NLFF modeling lowers the estimates of (and therefore possibly underestimates) flare-related changes, because the model magnetic fields are naturally more similar to each other than if they were based on %\so{higher resolution} 
data {at smaller pixels sizes} (due to the smoothing of possibly important magnetic flux and electric currents at small spatial scales).

More generally, the realism of any kind of modeling is usually assumed to increase if the truly involved spatial scales are accommodated adequately. In other words, modeling %\so{at higher spatial resolution} 
{involving smaller pixel sizes} is assumed to provide a better representation of the true complexity of the static magnetic corona. In Fig.~\ref{fig:fig8}, we visualize NLFF magnetic field models %\so{at different spatial resolutions} 
{with different plate scales}, using the example of
%\LEt{Please check that I have retained your intended meaning.} 
AR~12673 at 08:48 on 2017 September 6. Despite exhibiting some differences in morphology, the models at all three tested spatial resolutions reveal almost the same basic connectivity within the AR core, including the strongly twisted field along the solar north-south direction in the eastern part of the AR \citep[see also, e.g.,][for the visualization of morphological differences when using different free model parameters during optimization]{2019A&A...628A..50M}.  From case to case and based on a corresponding in-depth analysis of the magnetic field morphology, it therefore remains to be judged whether or not its inherent %\so{spatial resolution} 
{pixel size} is sufficient to provide model support for specific observed features.

Finally, to place all of the above into greater context, the overall changes induced by a change %\so{in spatial resolution} 
{of the spatial sampling} are small compared to those possibly induced by the use of different calibration products of a given instrument or the use of data from different instruments. For such cases, relative changes of the unsigned magnetic fluxes and magnetic energies by factors of $\gtrsim$2 were reported \citep{2012AJ....144...33T,2013ApJ...769...59T}. Also, differences arising from the application of different NLFF methods to the very same data set appear much larger, with method-induced differences of a factor of $\gtrsim$2 for free-energy estimates \citep{2015ApJ...811..107D}.

%%%%%%%%%%%%%%%%%%%%%%%%%%%%%%%%%%%%%%%%%%%%%%%%%%%%%%%%%%%%%%%%%%%%%%%%%%%%%%%%%%%%%
%%%%%%%%%%%%%%%%%%%%%%%%%%%%%%%%%%%%%%%%%%%%%%%%%%%%%%%%%%%%%%%%%%%%%%%%%%%%%%%%%%%%%
\section{Summary and Conclusion} \label{s:summary}
%%%%%%%%%%%%%%%%%%%%%%%%%%%%%%%%%%%%%%%%%%%%%%%%%%%%%%%%%%%%%%%%%%%%%%%%%%%%%%%%%%%%%
%%%%%%%%%%%%%%%%%%%%%%%%%%%%%%%%%%%%%%%%%%%%%%%%%%%%%%%%%%%%%%%%%%%%%%%%%%%%%%%%%%%%%

NLFF modeling is regularly used to indirectly infer the 3D geometry of the coronal magnetic field, which is not otherwise accessible on a regular basis by means of direct measurements. For such purposes, routinely measured photospheric magnetic field vector data binned to a %\so{coarser spatial resolution} 
{larger pixel size} are used as an input. However, this practice was suspected to affect the reliability of the modeling \citep{2009ApJ...696.1780D}. In a dedicated study that analyzes the resolution-dependence of different NLFF methods, \cite{2015ApJ...811..107D}  indeed demonstrated non-negligible effects. However, that work  was based on the analysis of NLFF modeling itself based on vector magnetic field data at a single time instant. It therefore remains unclear whether or not detected trends are to be expected in general. Moreover, it is difficult to compare %\so{spatial-resolution-induced} 
{spatial-sampling-induced} variations with method-induced ones, because different NLFF methods tested in  \cite{2015ApJ...811..107D}
 treat the input data very differently.

The aim of this work is to partially close those gaps. In order to study %\so{resolution-induced} 
{spatial-sampling-induced} effects systematically, we performed multi-snapshot NLFF modeling using a single NLFF \citep[optimization;][]{2012SoPh..281...37W} method. For three solar ARs (NOAAs 11158, 11429, and 12673), we used %\so{time-series} 
{time series} of SDO/HMI data at three different spatial resolutions: once at their native resolution, and reduced by factors of two and four. This allowed us to (1) study the effect of binning as a function of time (within %\so{time-series} 
{time series} of individual ARs), 2) spot very different %\so{resolution-induced}
{spatial-sampling-induced} changes for different ARs, and 3) deduce general trends.

Regarding items (1) and (2) above, we clearly demonstrate that a certain change %\so{in spatial resolution}  
{of the spatial sampling} does not necessarily translate to similar effects at another time instant within a %\so{time-series} 
{time series} of NLFF models for a particular AR, and also that the induced changes can be distinctly different for different ARs. This is true for both the magnitude of the induced changes as well as their ``direction'' (increasing or decreasing). From the detailed analysis of the %\so{different-resolution} 
HMI data-based NLFF model %\so{time-series} 
{time series with different underlying plate scales} of specifically chosen ARs, our findings are as follows.

\begin{enumerate}
        \item The overall success of NLFF modeling at a given %\so{spatial resolution} 
        {pixel size} (plate scale) is necessarily different for different ARs, but also varies considerably across the model %\so{time-series} 
        {time series} of individual ARs (Sect.~\ref{ss:nlff_quality}). Therefore, in agreement with past experience, concise quality checks are to be performed for every single NLFF model prior to any attempt to interpret deduced physical parameters.
        \item Among frequently used metrics to quantify the solenoidal quality of NLFF models, two measures deduced from magnetic energy decomposition appear most sensitive (thus indicative), namely the fraction of nonsolenoidal contributions to the total energy ($\Edivprime$) and the relative size of nonsolenoidal and free magnetic energy ($\Emixprime$). The recently proposed measure $\fdavg$ appears less sensitive in that respect.
        \item The solenoidal quality of a NLFF model neither relates to the underlying %\so{spatial resolution} 
        {spatial sampling} nor is to be found at similar levels for different ARs even when given the same underlying %\so{spatial resolution} 
        {plate scale} (Sect.~\ref{ss:nlff_quality} and Fig.~\ref{fig:fig2}).
        \item Binning of SDO/HMI data by a factor of four (to a plate scale of $\sim$1.44~Mm (our ``bin4'' case) may yield unphysical solutions (for which $\Efree<0$; see Sect.~\ref{sss:energies} and Fig.~\ref{fig:fig4}).
        \item The ultimate controlling parameter of the %\so{resolution-induced} 
        variations of the deduced physical quantities {induced by a changed plate scale} is the solenoidal quality of the NLFF model. This is evidenced by a  clear corresponding (1:1) relation of %\so{resolution-induced} 
        changes to amplitudes %\so{and} 
        {as well as} to temporal patterns. 
        %\textcolor[rgb]{0.984314,0.00784314,0.027451}{in parts clearly differing form those seen in the respective input data}.
	%\LEt{ please make this into a separate sentence as the meaning is not clear.}
	\item For each of our tested ARs, comparatively larger values of $\Etot$, $\Efree$, and $|\Hj|$, in conjunction with lower values of $\Epot$, $|\Hv|$, and $|\Hpj|$ (Sect.~\ref{sss:energies} and \ref{sss:helicities}) were found for models with lower values of $\Emixprime$ (and to a large degree also lower values of $\Edivprime$).
        \item Despite fluctuations, observed trends in the %\so{time-series} 
        {time series} of the model-deduced physical parameters for the individual ARs appear consistent across different resolutions, including, for example, phases of systematic increases or decreases and pronounced flare-related changes.
\end{enumerate}

Having our extended analysis at hand, we were also able to deduce some general trends for the application to HMI data (cf.\ Sect.~\ref{s:discussion}). Using $\tjavg$, $\Edivprime$ and $\Emixprime$ as measures, NLFF %\so{model} 
{modeling} quality tends to be higher %\so{at reduced spatial resolution} 
{for larger pixel sizes involved} (Fig.~\ref{fig:fig6}). Taken together with {simultaneously found} larger values of $\Etot$, %\so{we find that} 
$\Efree$ and $\Hj$ (Fig.~\ref{fig:fig7}) %\so{suggest} 
{suggests} that the optimization method %\so{may converge} 
{converges} to more satisfactory solutions. %\so{at lower spatial resolutions.}
%\LEt{Please check that I have retained your intended meaning; if not, please consider rewording the original text for improved clarity.}
{Importantly,} estimates of flare-related changes of $\Efree$ and $\Hj$ from NLFF modeling %\so{at lower resolution} 
{based on down-sampled input data} are found to be  systematically smaller, and are possibly underestimations of the true extent. Binning {(down-sampling)} of SDO/HMI data by a factor of two (to a plate scale of $\sim$0.72~Mm; ``bin2'') yields changes to the deduced volume-integrated {total and potential} magnetic energies %\so{, $\Epot$ and $\Etot$,} 
of $\lesssim$5\% and to the {total and volume-threading} relative helicities %\so{, $\Hv$ and $\Hpj$,} 
of $\lesssim$10\% (Fig.~\ref{fig:fig7}). %\so{However, for} 
{Though corresponding changes to} $\Efree$ and $\Hj$ %\so{they} 
are somewhat larger ($\approx$20\%), %\so{. Resolution-induced changes appear}  
{they are} relatively small compared to other possible sources of uncertainty, including effects related to the use of different calibration products, the use of input data from different instruments, or the application of different NLFF methods. {Finally,}  NLFF modeling %\so{at different spatial resolutions yields} 
%\so{at down-sampled spatial scales exhibits a} consistent %\so{results on the} 
{at down-sampled spatial scales appears to retain the} %\so{results on the} 
basic magnetic connectivity %\so{of an analyzed solar AR} 
(Fig.~\ref{fig:fig8}). %\so{The extent to which NLFF modeling at a specific spatial resolution is suited to providing} 
%\so{Whether or not suited to provide sufficient} model support for selected observational features remains to be demonstrated and evaluated independently for any employed model {in any ways}.

\begin{acknowledgements} 
We thank the anonymous referee for valuable suggestions to improve the clarity of the manuscript. J.K.T., M.G., and A.M.V. acknowledge Austrian Science Fund (FWF): P31413-N27. SDO data are courtesy of the NASA/SDO AIA and HMI science teams.
\end{acknowledgements}

%%%%%%%%%%%%%%%%%%%%%%%%%%%%%%%%%%%%%%%%%%%%%%%%%%%%%%%%%%%%%%%%%%%%%%%%%%%%%%%%%%%%%
%%%%%%%%%%%%%%%%%%%%%%%%%%%%%%%%%%%%%%%%%%%%%%%%%%%%%%%%%%%%%%%%%%%%%%%%%%%%%%%%%%%%%
\bibliographystyle{aa} % style aa.bst
\bibliography{bibliography} % your references Yourfile.bib
%%%%%%%%%%%%%%%%%%%%%%%%%%%%%%%%%%%%%%%%%%%%%%%%%%%%%%%%%%%%%%%%%%%%%%%%%%%%%%%%%%%%%
%%%%%%%%%%%%%%%%%%%%%%%%%%%%%%%%%%%%%%%%%%%%%%%%%%%%%%%%%%%%%%%%%%%%%%%%%%%%%%%%%%%%%

\end{document}